\documentclass[onecolumn]{arxiv}

\usepackage{amssymb}
\usepackage{amsmath}

\usepackage[sort,compress]{cite}

\usepackage{graphicx} 
\usepackage{epsfig} 

\newtheorem{proposition}[thm]{Proposition}
\newtheorem{corollary}[thm]{Corollary}
\newtheorem{theorem}[thm]{Theorem}
\newtheorem{remark}[thm]{Remark}
\newtheorem{lemma}[thm]{Lemma}
\newtheorem{definition}[thm]{Definition}
\newtheorem{assumption}[thm]{Assumption}

\newcommand*{\rom}[1]{\expandafter\@slowromancap\romannumeral #1@}

\usepackage{xcolor}

\usepackage{graphicx}          

\begin{document}

\begin{frontmatter}

\title{Convergence Tools for Consensus in Multi-Agent Systems with Switching Topologies\thanksref{footnoteinfo}}

\thanks[footnoteinfo]{The authors gratefully acknowledge the financial support
from the Swedish National Space Technology Research Programme (NRFP)
and the Swedish Research Council (VR).  
}

\author[Johan]{Johan Thunberg}\ead{johan.thunberg@math.kth.se},    
\author[Johan]{Xiaoming Hu}\ead{hu@math.kth.se}

\address[Johan]{KTH Royal Institute of Technology,  S-100 44 Stockholm, Sweden}


\begin{abstract}
We present two main theorems along the lines of Lyapunov's second method 
that guarantee asymptotic state consensus in multi-agent systems of agents in $\mathbb{R}^m$ with
switching interconnection topologies. The two theorems 
complement each other in the sense that the first one is formulated in terms of the states
of the agents in the multi-agent system, whereas the second one is formulated in terms of the 
pairwise states for each pair of agents in the multi-agent system.
In the first theorem, under the assumption that the interconnection topology is
uniformly strongly connected and that the agents
are contained in a compact set, a strong form of attractiveness of the consensus set
 is assured. In the second theorem, under the weaker assumption 
that the interconnection topology is uniformly quasi strongly connected,
the consensus set is guaranteed to be uniformly asymptotically stable.
\end{abstract}

\end{frontmatter}

\section{Introduction}
The field of networked and multi-agent systems has received growing
interest from researchers within robotics and control theory during the last
decade \cite{mesbahi2010graph}. This increased attention to network science is partly due to the
recent advancement of communication technologies such as cellular phones,
the Internet, GPS, wireless sensor networks etc. The widespread use and
ongoing development of such technologies is a testament to the great potential
applicability of the work carried out within this research field.

Consensus is a key problem in multi-agent systems theory and it has indeed also been 
one of the main objects of attention.
Early works include~\cite{reynolds1987flocks,vicsek1995novel}. Some of the most well cited publications within the 
control community are \cite{moreau2005stability,olfati2004consensus,olfati2007consensus,ren2005consensus}.
Due to the vast amount of publications, it is a challenge to provide
a complete overview of the subject, and in this introduction we merely provide a selection from the body 
of knowledge. There are books~\cite{mesbahi2010graph,ren2008distributed}, 
and surveys \cite{garin2010survey,ren2005survey} covering the subject from different perspectives.

The problem of consensus or state agreement can roughly be explained as follows. 
Given a multi-agent system where each agent has a state in a common space and
where the states are updated to a dynamical equation, design a distributed
control law for the system such that the states of the agents converge to the same 
value. The convergence is mostly defined in the asymptotic sense.

The dynamics for the agents can either be described in discrete time~\cite{montijano2011fast,xiao2007distributed}
or continuous time \cite{moreau2004stability}. 
This work considers
continuous time dynamics. Furthermore, if the dynamics is linear, much of the work  
has centered around graph theoretic
concepts such as the graph Laplacian matrix, and its importance for the convergence 
to the consensus manifold~\cite{mesbahi2010graph,murray2003consensus,olfati2004consensus}. 
For homogeneous systems of agents with linear dynamics, the question of which properties must hold in order to guarantee consensus has been answered~\cite{ma2010necessary}.

Here, similar to \cite{lin2007state,moreau2005stability,shi2009},
we consider a broad class of multi-agent systems and 
provide some criteria in order to guarantee consensus which are common for a class 
of systems. In those works, consensus is assured by imposing a convexity assumption.
 Roughly, provided the existence and uniqueness of solution is guaranteed,
if the right-hand side of each agent's dynamics,
 is inward-pointing~\cite{afsari2011riemannian} relative to
the convex hull of the position of itself and the positions of its neighbors (states), asymptotic consensus is guaranteed. 

Instead of using a convexity assumption, in this work we provide
two classes of functions. Provided certain conditions are fulfilled for the 
system and the functions, two theorems guarantee consensus or state agreement.
The first class of functions are functions of state, and the second
class of functions are functions of pairs of states.
The two theorems differ in the sense that the first one is formulated
for functions in the first class and the second one is formulated for functions in the
second class. The theorems can be combined in order to show consensus under the convexity assumptions
in  \cite{lin2007state,moreau2005stability,shi2009}.
However, as we show, there are examples when the convexity assumptions do not hold
but where the proposed theorems can be applied.

The proposed functions can be interpreted as Lyapunov-like functions in 
order to show consensus for multi-agent systems. If a function 
is used from the first class, a strong form of 
attractiveness of the consensus set is shown in the first theorem. On the contrary,
if a function from the second class is used,
 uniform asymptotic stability to the 
consensus manifold is shown in the second theorem. Even though the second theorem provides stronger 
conditions for convergence, the first theorem can in general be
applied in wider context.

We provide numerous examples of the usefulness of the theorems.
One such example regards nonlinear scaling in a well known
consensus control law for agents with single integrator dynamics. 
This control law consists of a weighted sum of the pairwise
differences between neighboring agents. 
In the modified nonlinear scaled version,
either the states have been scaled, or the differences between
the states have been scaled. If the differences have been scaled,
the control law falls into the frameworks of  \cite{lin2007state,moreau2005stability,shi2009}.
However, if the states are scaled, this situation is not captured by the 
convexity assumption, but the first theorem we present is still applicable.

Connectivity is key to achieving collective behavior
in a multi-agent system. In fact, the topologies for the practical
multi-agent networks may change over time. In the study of variable
topologies, a well-known connectivity assumption, called (uniform)
joint connection without requiring connectedness of the graph at
every moment, was employed to guarantee multi-agent consensus for
first-order or second-order linear or nonlinear systems
\cite{cheng2008,hong2007,jadbabaie2003,shi2009}. 

Under these mild switching conditions we allow
the right-hand side of the system dynamics to switch between
a finite set of functions that are piecewise continuous in 
the time and uniformly Lipschitz in the state on some compact region containing the 
origin. Similar to earlier works we assume a positive lower bound on the dwell time
 between two consecutive time instances where the right-hand side switches between two functions in 
the set. Also we require in general an upper bound on the dwell time
 (in the case of time invariant functions we do not require such an upper bound).

The time dependence in the right-hand side of the system dynamics is restricted in the sense
that it only depends on the time since the last switch between two functions. 
This type of time dependence can be used in a wide range of applications,
for example one can show that for a system switching between a finite set of time invariant functions,
one can define continuous in time transitions between the functions instead of discontinuous switches, so that the
right-hand side of the system dynamics is continuous and the system reaches consensus with the 
same rate of convergence as for the switching system.

\section{Preliminaries}\label{sec:preliminaries}
\subsection{Dynamics}\label{sec:dynamics}
Let us introduce the following finite set of functions 
$$\mathcal{F}= \{\tilde{f}_1(t,x), \ldots, \tilde{f}_{|\mathcal{F}|}(t,x)\},$$
where 
$$\tilde{f}_k: \mathbb{R} \times \mathbb{R}^{mn} \rightarrow \mathbb{R}^{mn}, \text{ for all } k = \{1, \ldots,|\mathcal{F}|\}, $$
is continuous in $t$ and Lipschitz in $x$, uniformly with respect to $t$,
on some open connected set containing 
the compact region $\mathcal{D} \in \mathbb{R}^{mn}$. We assume that $\mathcal{D}$ contains the origin as an interior point.
 The symbol 
$|\mathcal{F}|$ is the number of functions in $\mathcal{F}$. 
Each function $\tilde{f}_k \in \mathcal{F}$ can be
written as $\tilde{f}_k = (\tilde{f}_{k,1}, \ldots, \tilde{f}_{k,n})^T$, where 
$$\tilde{f}_{k,l}: \mathbb{R} \times \mathbb{R}^{mn} \rightarrow \mathbb{R}^m \quad \text{ for all } \: l.$$ 

By following \cite{cheng2008}, we define  
switching signal functions which will be used in the 
definition of the system dynamics.
We will assume that a switching signal function $\sigma$ satisfies either Assumption~\ref{chapter1:sigma_assumption} (1,2)
 or Assumption~\ref{chapter1:sigma_assumption} (1,2,3) below (what we mean by
\emph{e.g.}, (1,2) is that the conditions 1 and 2 are satisfied). 
\begin{assumption}\label{chapter1:sigma_assumption}
~
\begin{enumerate}
\item The function $\sigma(t): \mathbb{R} \rightarrow \{1, \ldots, |\mathcal{F}|\}$ is piecewise right-continuous. \\
\item There is a monotonically increasing sequence $\{\tau_k\}$, such that $\tau_k \rightarrow \infty$ as $k \rightarrow \infty$
and $\tau_k \rightarrow -\infty$ as $k \rightarrow -\infty$, 
where each $\tau_k \in \mathbb{R}$ is such that for any $k \in \mathbb{Z}$
the function $\sigma$ is constant on $[\tau_{k}, \tau_{k+1})$ for all $k$, and there is a  $\tau_D > 0$
such that 
\begin{equation*}
\inf_k (\tau_{k+1}-\tau_k) \geq \tau_D \text{ and }
\end{equation*}

\item there is an upper bound $\tau_U > 0$, such that for any 
\begin{equation*}
\sup_k (\tau_{k+1}-\tau_k) \leq \tau_U. \quad 
\end{equation*}
\end{enumerate}
\end{assumption}

We define the set of all functions $\sigma$ that 
 fulfills Assumption~\ref{chapter1:sigma_assumption} (1,2) as $\mathcal{S}_{|\mathcal{F}|,D}$ and 
fulfills Assumption~\ref{chapter1:sigma_assumption} (1,2,3) as $\mathcal{S}_{|\mathcal{F}|,D,U}$. 
The constants $\tau_D$ and $\tau_U$ might be different for different $\sigma$,
so condition 2 and 3 in Assumption~\ref{chapter1:sigma_assumption} can also be formulated as
$$\inf_k(\tau_{k+1} - \tau_k) > 0 \quad \text{ and } \quad  \sup_k(\tau_{k+1} - \tau_k) < \infty$$
respectively.
For each $\sigma$, the sequence $\{\tau_k\}$ is referred to as the switching 
times of $\sigma$, since it is only at those times $\sigma(t)$ changes value.
If we compare the upper and lower bounds for two switching signal functions $\sigma_1$ and
$\sigma_2$, we denote the upper and lower bound for $\sigma_1$ as 
$\tau_U^{\sigma_1}$ and $\tau_D^{\sigma_1}$ respectively and the upper 
and lower bound for $\sigma_2$ as $\tau_U^{\sigma_2}$ and $\tau_D^{\sigma_2}$ respectively.

For a given $\sigma \in \mathcal{S}_{|\mathcal{F}|,D}$ with 
switching times $\{\tau_k\}$ we define (for finite times)
$$\gamma_{\sigma}(t) = \max\{\tau_k :\tau_k \leq t, k \in \mathbb{Z} \},$$
where $\gamma_{\sigma}(t)$ is the largest switching time less than or equal to $t$.

Let us now consider a system of $n$ agents. The state of agent
$i$ at time $t$ is defined as $x_i(t) \in \mathbb{R}^m$. The dynamics 
for the system of agents that we consider is given by
\begin{align*}
\dot{x}_1  = &~f_1(t,x) = \tilde{f}_{\sigma(t),1}(t - \gamma_{\sigma}(t),x),\\
 \vdots  \:&\\
\dot{x}_n = &~f_n(t,x) = \tilde{f}_{\sigma(t),n}(t - \gamma_{\sigma}(t),x),
\end{align*}
where $\sigma \in \mathcal{S}_{|\mathcal{F}|,D}$ and 
$$(\tilde{f}_{\sigma(t),1}, \ldots, \tilde{f}_{\sigma(t),n})^T = \tilde{f}_{\sigma(t)} \in \mathcal{F}.$$
Note that $f_i(t,x) \in \mathbb{R}^m$ for $i \in \{1, \ldots, n\}$, whereas $\tilde{f}_i(t,x) \in
\mathbb{R}^{mn}$ for $i \in \{1, \ldots, |\mathcal{F}|\}$.
The main results 
in this work regard the restricted case when $\sigma \in \mathcal{S}_{|\mathcal{F}|,D,U}$,
however there are cases when we assume the general case when $\sigma \in \mathcal{S}_{|\mathcal{F}|,D}$.
The system dynamics can be written as
\begin{equation}\label{chapter1:dynamics}
\dot{x} = f(t,x) = \tilde{f}_{\sigma(t)}(t - \gamma_{\sigma}(t),x),
\end{equation}
where,
$f(t,x) = \: (f_1(t,x), \ldots, f_n(t,x))^T.$
For a given $\sigma$, the function $f(t,x)$ is piecewise continuous in $t$. It is Lipschitz in $x$ on
$\mathcal{D}$, uniformly with respect to $t$.
The initial state and the initial time for \eqref{chapter1:dynamics} is 
$x_0 \in \mathcal{D}$ and $t_0$ respectively. Sometimes we write $x(t_0)$ instead of $x_0$. 

The switching signal functions are used in order
to indicate which system we are referring to.
For a given $\mathcal{F}$,
the switching behavior of the system is  
captured by $\sigma$. In order to emphasize this, instead of writing $x$ 
we can write
$$x^{\sigma} = (x^{\sigma}_1, \ldots, x^{\sigma}_n)^T.$$
In general we omit the parametrization by $\sigma$ and 
write $x$ instead of $x^{\sigma}$, but the latter notation is useful
when we study solutions of \eqref{chapter1:dynamics} for different choices of $\sigma$.
The solution for the system \eqref{chapter1:dynamics} is sometimes also written as
$x(t,t_0,x_0)$ or $x^{\sigma}(t,t_0,x_0)$, where the explicit dependence on the initial time $t_0$ and
the initial state $x_0$ is emphasized.

\begin{lemma}\label{lem:a1}
If all the functions in $\mathcal{F}$ are 
time-invariant, the dynamics \eqref{chapter1:dynamics} is given by 
$$\dot{x} = \tilde{f}_{\sigma(t)}(x),$$
and if $\sigma \in \mathcal{S}_{|\mathcal{F}|,D}$ but $\sigma \not\in \mathcal{S}_{|\mathcal{F}|,D,U}$, it holds that there is a corresponding $\sigma' \in \mathcal{S}_{|\mathcal{F}|,D,U}$
for which the dynamics is the same. {i.e.}, 
$$\dot{x} = \tilde{f}_{\sigma'(t)}(x) =  \tilde{f}_{\sigma(t)}(x)$$
for all $t \geq 0$.
\end{lemma}

\begin{lemma}\label{lem:a2}
For $\sigma \in \mathcal{S}_{|\mathcal{F}|,D,U}$
with lower bound $\tau_D^{\sigma}$ and upper bound $\tau_U^{\sigma}$ on the dwell time between two consecutive switches, 
there is a finite set of functions (continuous in $t$ and Lipschitz in $x$ on $\mathcal{D}$, uniformly with respect to $t$)
$$\mathcal{F}' = \{\tilde{f}_1', \ldots, \tilde{f}_{|\mathcal{F}'|}'\} \supset \mathcal{F}$$ 
and $\sigma' \in \mathcal{S}_{|\mathcal{F}'|,D,U}$ with a lower bound $\tau_D^{\sigma'} = \tau_D^{\sigma}$ and an upper bound
 $\tau_U^{\sigma'} = 2\tau_D^{\sigma}$ on the dwell time
between two consecutive switches,
such that $$\tilde{f}'_{\sigma'(t)}(t- \gamma_{\sigma'}(t),x) = \tilde{f}_{\sigma(t)}(t- \gamma_{\sigma}(t),x).$$
\end{lemma}
The proofs of these lemmas as well as all other proofs that are not given directly
are contained in Section~\ref{sec:proofs}. 
Due to Lemma~\ref{lem:a2}, 
we will often
consider the case when $\tau_U = 2\tau_D$ since we can replace $\mathcal{F}$ with $\mathcal{F}'$ and $\sigma$ with $\sigma'$.
Note that $\tau_D^{\sigma}$ and $\tau_U^{\sigma}$ do not need to be the greatest lower bound and the least upper bound respectively for the 
dwell time between two consecutive switches of $\sigma$.

\subsection{Connectivity}\label{sec:con}
In a multi-agent
system the dynamical behavior in general depends on the
connectivity between the agents. The connectivity is described
by a graph.

\begin{definition}\label{def:chapter1:graff}
A directed graph (or digraph) $\mathcal{G} = (\mathcal{V},
\mathcal{E})$ consists of a set of nodes,  $\mathcal{V} = \{1, ..., n\}$ 
and a set of edges $\mathcal{E} \subset \mathcal{V} \times \mathcal{V}$.
\end{definition}
In our setting, each node in the graph corresponds to a unique agent.
Thus $\mathcal{V}$ is henceforth defined
as $\mathcal{V} = \{1, ..., n\}$.
We also define neighbor sets or neighborhoods.
Let $\mathcal{N}_i \in
\mathcal{V}$ comprise the neighbor set (sometimes referred to simply as neighbors)
of agent $i$, where $j \in \mathcal{N}_i$
if and only if $(j,i) \in \mathcal{E}$. We assume throughout the thesis that $i \in
\mathcal{N}_i$ \emph{i.e.}, we restrict the collection of graphs
to those for which $(i,i) \in \mathcal{E}$ for all $i \in \mathcal{V}$. 

A directed path of $\mathcal{G}$ is an ordered sequence of distinct nodes in $\mathcal{V}$
such that any consecutive pair of nodes in the sequence corresponds to an edge in the graph.
An agent $i$ is connected to an agent $j$ if there is a directed path starting in $j$ and ending in $i$.

\begin{definition}
A digraph is strongly connected if each node $i$
is connected to all other nodes.
\end{definition} 
\begin{definition}
A digraph is quasi-strongly connected if there exists a rooted 
spanning tree or a center, \emph{i.e.}, at least one node such
that all the other nodes are connected to it.
\end{definition} 

We are now ready to address time-varying graphs. 
From Definition~\ref{def:chapter1:graff} we see that there are
$2^{n^2}$ possible directed graphs with $n$ nodes.
For $k \in \{1, \ldots, |\mathcal{F}|\}$ we associate a corresponding graph $\mathcal{G}_k = (\mathcal{V}, \mathcal{E}_{k})$. 
Note that the graphs $\mathcal{G}_k$ and
$\mathcal{G}_l$ might be the same for $k \neq l$ (\emph{i.e.},
the set of edges is equal for the two graphs $\mathcal{G}_k$ and $\mathcal{G}_l$).

For $\sigma \in \mathcal{S}_{|\mathcal{F}|,D}$ we define the time-varying graph corresponding to $\sigma$
as $\mathcal{G}_{\sigma(t)}$
and the time-varying neighborhoods as
$\mathcal{N}_i(t)$ for all $i$. If we want to emphasize explicitly which switching signal function is used,
we write $\mathcal{N}_i^{\sigma}(t)$ or $\mathcal{N}_i^{\sigma(t)}$. 

\begin{definition}
For $\sigma \in \mathcal{S}_{|\mathcal{F}|,D}$,
the union graph of $\mathcal{G}_{\sigma(t)}$ during the time
interval $[t_1,t_2)$ is defined as
\begin{equation*}
\mathcal{G}([t_1, t_2))
= \textstyle\bigcup_{t\in[t_1, t_2)} \mathcal{G}_{\sigma(t)}
= (\mathcal{V},\textstyle\bigcup\nolimits_{t\in[t_1, t_2)}\mathcal{E}_{\sigma(t)}),
\end{equation*}
where $t_1 < t_2 \leq +\infty$.
\end{definition}
\begin{definition}\label{def:quasi}
The graph $\mathcal{G}_{\sigma(t)}$ is uniformly
(quasi-) strongly connected if $\sigma \in \mathcal{S}_{|\mathcal{F}|,D}$ and there exists a constant $T^{\sigma}>0$ such that the
union graph $\mathcal{G}([t, t + T^{\sigma}))$ is (quasi-) strongly connected for all $t$.
\end{definition}

\subsection{Some special functions, sets and operators}
\begin{definition}\label{def:v1}
For
$V:\mathbb{R}^m \rightarrow \mathbb{R}$
we define $f_{V,m}: \mathbb{R}^{mn} \rightarrow \mathbb{R}$ as 
$$f_{V,m}(x) =  \max_{j \in \mathcal{V}} V(x_j).$$
\end{definition}
\begin{definition}\label{def:v1}
For $W: \mathbb{R}^m \times \mathbb{R}^m \rightarrow \mathbb{R}$
we define ${f}_{W, m,m}: \mathbb{R}^{mn} \rightarrow \mathbb{R}$ as 
$${f}_{W,m,m}(x) =  \max_{(i,j) \in \mathcal{W} \times \mathcal{V}} V(x_i,x_j).$$
\end{definition}

\begin{definition}
Suppose for $\sigma \in \mathcal{S}_{|\mathcal{F}|,D}$ that $x^{\sigma}$ is a solution to \eqref{chapter1:dynamics}
and $x^{\sigma}(t)$ is contained in $\mathcal{D}$ on an interval $[t_0, t_0 + \tilde{t})$ where $\tilde{t} > 0$. Suppose also that $V:\mathbb{R}^m \rightarrow \mathbb{R}$
and $W: \mathbb{R}^m \times \mathbb{R}^m \rightarrow \mathbb{R}$ are continuously differentiable.
On $[t_0, t_0 + \tilde{t})$, let
\begin{align*}
\mathcal{I}_V(t_1,t_2) & = \left\{i: V(x_i(t_2)) = f_{V,m}(x(t_1))\right\}, \\
\mathcal{J}_W(t_1,t_2) & = \left\{(i,j): W(x_i(t_2), x_j(t_2)) = f_{W,m,m}(x(t_1))\right\}, \\
\mathcal{I}^*_V(t) & = \mathcal{I}_V(t,t) \cap \left\{i: \frac{d}{dt}{V}(x_i(t)) < 0, i \in \mathcal{V} \right\}, \\
\mathcal{J}^*_W(t) & = \mathcal{J}_W(t,t) \cap \left\{(i,j): \frac{d}{dt}{W}(x_i(t), x_j(t)) < 0, (i,j) \in \mathcal{V} \times \mathcal{V}\right\}.
\end{align*}
\end{definition}
These sets, except for being functions of the times $t_1, t_2$ or $t$, also depend on the initial 
conditions $x_0$, $t_0$ and the switching signal function. 
In order
to simplify the notation, we do not parameterize 
these sets by $\sigma$, $t_0$ and $x_0$.

The upper Dini derivative of a function $V(t,x(t))$ with respect to $t$ is defined as 
$$D^+V(t,x(t)) = \limsup_{\epsilon \downarrow 0}\frac{V(t + \epsilon, x(t + \epsilon)) - V(t, x(t))}{\epsilon}.$$
Given this definition we now proceed with a useful lemma, \cite{shi2009,lin2007state}.
\begin{lemma}\label{lem:dini}
~

\begin{itemize}
\item If $V:\mathbb{R}^m \rightarrow \mathbb{R}$ is continuously differentiable, then 
$$D^+f_{V,m}(x(t)) = \max_{i \in \mathcal{I}_{V}(t,t)}\frac{d}{dt}{V}(x_i(t)).\quad \quad \quad \:\: \:$$ 
\item If $V:\mathbb{R}^m \times \mathbb{R}^m \rightarrow \mathbb{R}$ is
continuously differentiable, then 
$$D^+f_{V,m,m}(x(t)) = \max_{(i,j) \in \mathcal{J}_{V}(t,t)}\frac{d}{dt}{V}(x_i(t),x_j(t)).$$
\end{itemize}
\end{lemma}

\subsection{Stability}
Let us introduce two equivalent definitions of uniform stability for the origin of
\eqref{chapter1:dynamics}. The first one is similar to the classic version~\cite{khalil2002nonlinear},
whereas the second one is a multi-agent systems version. In
the definitions of stability here, we consider
the stability for a set or a family of systems, where the systems in the set differ in the choice of switching signal
function $\sigma$. Thus, the stability holds for all choices of switching signal
functions in $\mathcal{S}_{|\mathcal{F}|,D}$, where the right-hand side of \eqref{chapter1:dynamics}
switches between functions in $\mathcal{F}$.

We assume that all the balls in the following definition are contained in $\mathcal{D}$.
The existence of such regions is assured by the assumption that $0$ is in the interior of $\mathcal{D}$.
\begin{definition}\label{stable:1}
~
\begin{enumerate}
\item The point $0 \in \mathbb{R}^{mn}$ is uniformly stable for \eqref{chapter1:dynamics} if
for $\varepsilon > 0$, there is $\delta(\epsilon) > 0$ such that
\begin{align*}
& x^{\sigma}(t_0) \in \bar{B}_{\delta, mn} \Rightarrow x^{\sigma}(t) \in \bar{B}_{\epsilon, mn},  \quad \text{for all } t \geq t_0, \sigma \in \mathcal{S}_{|\mathcal{F}|,D}.
\end{align*}
\item The point $0 \in \mathbb{R}^m$ is uniformly stable for \eqref{chapter1:dynamics} if
for $\varepsilon > 0$, there is $\delta(\epsilon) > 0$ such that
\begin{align*}
& x^{\sigma}_i(t_0) \in \bar{B}_{\delta, m} \Rightarrow x^{\sigma}_i(t) \in \bar{B}_{\epsilon, m},  \quad \text{for all } i, t\geq t_0, \sigma \in \mathcal{S}_{|\mathcal{F}|,D}.\\
\end{align*}
\end{enumerate}
\end{definition}
In the multi-agent systems
setting it feels often more intuitive to define the region of stability in the space where
the agents reside, using 2, since then each agent only needs to check that its state is inside the region of stability.

For a set $\mathcal{A} \subset \mathbb{R}^{mn}$, let $$\text{dist}(x,\mathcal{A}) = \inf_{y \in \mathcal{A}}\|x - y\|. $$
We say that $x(t)$ approaches $\mathcal{A}$ or $x(t) \rightarrow \mathcal{A}$ as $t \rightarrow \infty$, on a subset of $\mathcal{D}$
if for all $\epsilon > 0$ and $x_0$ in the subset, there exists $T(\epsilon,x_0,t_0)$ such that $\text{dist}(x(t),\mathcal{A}) < \epsilon$
for all $t \geq T(\epsilon,x_0,t_0) + t_0$. 
Let us proceed with the definition of invariance 
of a set for the system \eqref{chapter1:dynamics}.
We start with the standard definition of invariance,
and proceed with the multi-agent systems definition
which is similar to the one in \emph{e.g.},~\cite{lin2007state}. 

\begin{definition} 
~
\begin{enumerate}
\item 
A set $\mathcal{A} \in \mathbb{R}^{mn}$ 
is (positively) invariant for the system \eqref{chapter1:dynamics} if for all $t_0$, it holds
that $$x_0 \in \mathcal{A} \Longrightarrow x^{\sigma}(t,t_0,x_0) \in \mathcal{A}$$ for all $t > t_0$ and $\sigma \in \mathcal{S}_{|\mathcal{F}|,D}$. \\
\item 
A set $\mathcal{A} \in \mathbb{R}^m$ 
is (positively) invariant for the system \eqref{chapter1:dynamics} if for all $i, t_0$, it holds
that $$ x^{\sigma}_i(t_0) \in \mathcal{A} \Longrightarrow x^{\sigma}_i(t,t_0,x_0) \in \mathcal{A}$$ for all $i$, $t > t_0$ and $\sigma \in \mathcal{S}_{|\mathcal{F}|,D}$.
\end{enumerate}
\end{definition}
When we use either one of these definitions, the choice should be apparent by the context. We define
\begin{align*}
 \mathcal{D}^*(\tilde{t}) 
= & \{x_0 \in \mathbb{R}^{mn}:  x^{\sigma}(t,t_0,x_0) \in \mathcal{D} \: \: \text{ for all } t_0, t \in [t_0, t_0+ \tilde{t}) \text{ and }  \sigma \in \mathcal{S}_{|\mathcal{F}|,D}\}
\end{align*}
and formulate the following lemma.
\begin{lemma}\label{lem:79}
For any $\tilde{t} \in [0,\infty]$, the set $\mathcal{D}^{*}(\tilde{t})$ is compact and
the set $\mathcal{D}^{*}(\infty)$ is also invariant.
\end{lemma}

In the definitions
of stability of the origin and the definitions of invariance, we assumed that $\sigma \in \mathcal{S}_{|\mathcal{F}|,D}$ is arbitrary, \emph{i.e.},
the statements must hold for any $\sigma \in  \mathcal{S}_{|\mathcal{F}|,D}$. However,
in the definitions of stability of a set which we now are to formulate, we only consider the case when $\sigma$ is 
fixed. Thus, in the following definitions we write $x$ instead of $x^{\sigma}$.
We restrict the state to be contained in the invariant compact set
 $\mathcal{D}^*(\infty)$. Hence, the stability of the set is only defined in the 
relative sense, relative to $\mathcal{D}^*(\infty)$. 
In these definitions we assume that $\mathcal{D}^*(\infty)$ is nonempty,
and we will later show how to assure this.

\begin{definition}\label{def:stability}
For \eqref{chapter1:dynamics} where $\sigma \in \mathcal{S}_{|\mathcal{F}|,D}$, the set $\mathcal{A}$ is
\begin{enumerate}
\item stable relative to $\mathcal{D}^*(\infty)$ if
for all $t_0$ and 
for all $ \epsilon >0$, there is $\delta( t_0, \epsilon) >0$ such that for $x_0 \in \mathcal{D}^*(\infty)$ it holds that
$$\textnormal{dist}(x_0,\mathcal{A})  \leq \delta \Longrightarrow \textnormal{dist}(x(t),\mathcal{A}) \leq \epsilon \quad \textnormal{for all } t \geq t_0.$$
\item uniformly stable relative to $\mathcal{D}^*(\infty)$ if it fulfills 1
and $\delta$ as a function of $t_0$ is constant.

\item attractive relative to $\mathcal{D}^*(\infty)$ if there is $c(t_0)$ such 
that $x(t) \rightarrow \mathcal{A}$ as $t \rightarrow \infty$ for all $x_0 \in \mathcal{D}^*(\infty)$ such that $\textnormal{dist}(x_0,\mathcal{A})  \leq c$. 

\item uniformly attractive relative to $\mathcal{D}^*(\infty)$ if it fulfills 3 and
$c$ as a function of $t_0$ is constant. Furthermore, if $\text{dist}(x_0, \mathcal{A}) \leq c$, for $ \eta >0$ there is $T(\eta)$ such
that $$t \geq t_0 + T(\eta) \Longrightarrow \textnormal{dist}(x(t), \mathcal{A}) < \eta.$$

\item asymptotically stable relative to $\mathcal{D}^*(\infty)$ if it fulfills 1 and 3. 

\item uniformly asymptotically stable relative to $\mathcal{D}^*(\infty)$ if it fulfills 2 and 4. 

\item globally uniformly asymptotically stable relative to $\mathcal{D}^*(\infty)$, if it fulfills 6 and 
$$c = \sup_{y \in \mathcal{D}^*(\infty)}\text{dist}(y,\mathcal{A}).$$

\item globally quasi-uniformly attractive relative to $\mathcal{D}^*(\infty)$
if $x(t) \rightarrow \mathcal{A}$ as $t \rightarrow \infty$ for all $x_0 \in \mathcal{D}^*(\infty)$ and all $t_0$. Furthermore,
 for all $ \eta >0$ there is $T(\eta)$ such
that $$\min_{t \in [t_0, t_0 + T(\eta)]}\textnormal{dist}(x(t), \mathcal{A}) < \eta$$
for all $x_0 \in \mathcal{D}^*(\infty)$ and $t_0$.
\end{enumerate}
\end{definition}

Let us in the following choose the set $\mathcal{A}$ as the consensus set,
\emph{i.e.}, 
$$\mathcal{A} = \{x = (x_1, \ldots, x_n)^T \in \mathbb{R}^{mn}: x_i = x_j \: \textnormal{for all } i,j\}.$$

We now formulate an assumption that creates a relationship between the functions in $\mathcal{F}$
and the neighborhoods of the agents. 

\begin{assumption}\label{ass:main:0}
For any given $t$ and $\sigma \in \mathcal{S}_{|\mathcal{F}|,D}$,
it holds that 
$\tilde{f}_{\sigma(t),i}(s,x)$ is, except for being a function of $s$, only a function of $\{x_j\}_{j \in \mathcal{N}^{\sigma(t)}_i}$ for all $s$,  $i \in \mathcal{V}$,
and $x \in \mathcal{D}$.
\\ Or equivalently.
$\tilde{f}_{k,i}(s,x)$ is, except for being a function of $s$, only a function of $\{x_j\}_{j \in \mathcal{N}^{k}_i}$ for all $s$,  $i \in \mathcal{V}$, $x \in \mathcal{D}$
and $k \in \{1, \ldots, |\mathcal{F}|\}$.
\end{assumption}

We continue with two central assumptions.

\begin{assumption}\label{ass:main:1}
Let $V:\mathbb{R}^m \rightarrow \mathbb{R}$ be a continuously differentiable function on $\mathcal{D}$.
 The function $V$ fulfills the following.
\begin{enumerate}
\item V is positive definite. \\
\item For any initial time $t_0$, initial state $x_0 \in \mathcal{D}$ and $\sigma \in \mathcal{S}_{|\mathcal{F}|,D}$, if there
is $\epsilon > 0$ such that the solution to \eqref{chapter1:dynamics} exists and is contained in 
$\mathcal{D}$ during $[t_0, t_0 + \epsilon)$, then for $t \in [t_0, t_0 + \epsilon)$ it holds that 
$$D^+f_{V,m}(x^{\sigma}(t)) \leq 0 \quad \text{ and }$$ 
\item for each agent $i \in \mathcal{I}_V(t,t)$ it holds that
$i \in \mathcal{I}_V^*(t)$ if there is $j \in \mathcal{N}^{\sigma}_i(t)$
such that $x^{\sigma}_i(t) \neq x^{\sigma}_j(t)$. Furthermore, if $i \in \mathcal{I}_V(t,t)$ and $i \not\in \mathcal{I}_V^*(t)$
it holds that $\tilde{f}_{\sigma(t), i}(s,x) = 0$ for all $s$.
\end{enumerate}
\end{assumption}

\begin{assumption}\label{ass:main:2}
Let $V:\mathbb{R}^m \times \mathbb{R}^m \rightarrow \mathbb{R}^+,$
be a continuously differentiable on $\mathcal{D}$. 
The function $V$ fulfills the following.
\begin{enumerate}
\item $V(x, y) = 0$ if and only if $x = y$, \\
\item  For any initial time $t_0$, initial point $x_0 \in \mathcal{D}$ and $\sigma \in \mathcal{S}_{|\mathcal{F}|,D}$, if there
is an $\epsilon > 0$ such that the solution to \eqref{chapter1:dynamics} exists and is contained in 
$\mathcal{D}$ during $[t_0, t_0 + \epsilon)$, then for $t \in [t_0, t_0 + \epsilon)$
$$D^+f_{V,m,m}(x^{\sigma}(t)) \leq 0\quad \text{and }$$
\item for each pair of agents $(i,j) \in \mathcal{J}_V(t,t)$ it holds that
$(i,j) \in \mathcal{J}_V^*(t)$ if there is $k \in \mathcal{N}^{\sigma}_i(t)$
such that $x^{\sigma}_i(t) \neq x^{\sigma}_k(t)$, or there is $l \in \mathcal{N}^{\sigma}_j(t)$
such that $x^{\sigma}_j(t) \neq x^{\sigma}_l(t)$. Furthermore, if $(i,j) \in \mathcal{J}_V(t,t)$ and $(i,j) \not\in \mathcal{J}_V^*(t)$
it holds that $\tilde{f}_{\sigma(t), i}(s,x) = 0$ and $\tilde{f}_{\sigma(t), j}(s,x) = 0$ for all $s$, and
\item for each pair of agents $(i,j) \in \mathcal{J}_V(t,t)$ it holds that
$(i,j) \in \mathcal{J}_V^*(t)$ only if there is $k \in \mathcal{N}^{\sigma}_i(t)$
such that $x^{\sigma}_i(t) \neq x^{\sigma}_k(t)$, or there is $l \in \mathcal{N}^{\sigma}_j(t)$
such that $x^{\sigma}_j(t) \neq x^{\sigma}_l(t)$.
\end{enumerate}
\end{assumption}
The easiest way to verify that 2-3 are fulfilled in 
Assumption~\ref{ass:main:1} and 2-4 are fulfilled in Assumption~\ref{ass:main:2}, is to use
Lemma~\ref{lem:dini}. For example the condition  2 in
Assumption~\ref{ass:main:1} can be verified as follows. If $x \in \mathcal{D}$ and 
$$i = \arg\max_{k \in \mathcal{V}}(V(x_k)),$$
where $x = (x_1, \ldots, x_n) ^T$, then if $\nabla V(x_i)\tilde{f}_j(t,x) \leq 0$
for all $j \in \{1, \ldots, |\mathcal{F}_i|\}$, 2 is fulfilled.
Condition 2 in Assumption~\ref{ass:main:2} is verified in the 
analogous way.

\section{Main results}\label{sec:main}

\begin{theorem}\label{chapter1:thm:1}
Suppose Assumption \ref{ass:main:1} (1,2) holds, then
$0$ is uniformly stable for \eqref{chapter1:dynamics}.
Furthermore, suppose that $\widehat{\beta}_1$ and $\widehat{\beta}_2$ are 
class $\mathcal{K}$ functions such that 
$$\widehat{\beta}_1(\|y\|) \leq V(y) \leq \widehat{\beta}_2(\|y\|),$$
then for $\epsilon$ such that $(\bar{B}_{\epsilon,m})^n \subset \mathcal{D}$, it holds that
$$x_0 \in \bar{B}_{\delta, m} \Longrightarrow x_i^{\sigma}(t,t_0,x_0) \in \bar{B}_{\epsilon, m}, \quad \text{ for all } i,t\geq t_0, \sigma 
\in \mathcal{S}_{|\mathcal{F}|,D},$$
where $\delta = \widehat{\beta}_2^{-1}(\widehat{\beta}_1(\epsilon))$.
\end{theorem}

\begin{theorem}\label{chapter1:thm:3}
Suppose assumptions \ref{ass:main:0} and \ref{ass:main:1} (2,3) hold and  $\sigma \in \mathcal{S}_{|\mathcal{F}|,D,U}$
is such that 
$\mathcal{G}_{\sigma(t)}$ is uniformly strongly connected,
then the consensus set $\mathcal{A}$ is globally quasi-uniformly attractive relative to $\mathcal{D}^*(\infty)$.
\end{theorem}

\begin{theorem}\label{chapter1:thm:5}
Suppose
assumptions \ref{ass:main:0} and \ref{ass:main:2} hold,
and $\sigma \in \mathcal{S}_{|\mathcal{F}|,D,U}$. It follows that the consensus set
$\mathcal{A}$ is globally uniformly asymptotically stable relative to $\mathcal{D}^*(\infty)$
if and only if $\mathcal{G}_{\sigma(t)}$ is uniformly quasi-strongly connected.
\end{theorem}

\begin{remark}
What we mean when we say that Assumption \ref{ass:main:1} (2,3) hold, is that
everything in Assumption \ref{ass:main:1} holds except possibly {(1)}.
This notation will be used throughout the chapter.
\end{remark}

\begin{remark}
If Assumption~\ref{ass:main:0} holds and Assumption~\ref{ass:main:2} (1,2,3) holds, Theorem~\ref{chapter1:thm:5} holds provided that 
the phrase "{if and only if $\mathcal{G}_{\sigma(t)}$ is uniformly quasi-strongly connected}"
is replaced with "{if $\mathcal{G}_{\sigma(t)}$ is uniformly quasi-strongly connected}".
\end{remark}

\begin{remark}
 Provided 
Assumption \ref{ass:main:1} (1,2) hold,
we can show that $\mathcal{D}^*(\infty)$ is nonempty, and
an easy way of guaranteeing that $x_0 \in \mathcal{D}^*(\infty)$ is
to use Theorem~\ref{chapter1:thm:1} and let $x_0 \in (\bar{B}_{\delta,m})^n \subset \mathcal{D}^*(\infty)$.
When we know that $\mathcal{D}^*(\infty)$ is nonempty and $x_0 \in \mathcal{D}^*(\infty)$,
we do not require $V$ to be positive definite in Theorem~\ref{chapter1:thm:3}, \emph{i.e.},
it is sufficient that only conditions 2 and 3 hold for $V$ in Assumption \ref{ass:main:1}.
This means that we can use one positive definite function $V_1$ in Theorem~\ref{chapter1:thm:1} in order
to construct a set that is contained in $\mathcal{D}^*(\infty)$, and another not necessarily positive definite function $V_2$,
in order to show that $\mathcal{A}$ is attractive in Theorem~\ref{chapter1:thm:3}. 
\end{remark}

We proceed with two corollaries. These corollaries follow as 
a consequence of the fact that if the functions in $\mathcal{F}$ are 
time-invariant and $\sigma \in \mathcal{S}_{|\mathcal{F}|,D}$, then there
is $\sigma'$ in $\mathcal{S}_{|\mathcal{F}|,D,U}$ such that 
$f_{\sigma(t)}(x) = f_{\sigma'(t)}(x)$ for all $t \geq 0$, see Lemma~\ref{lem:a1}.

\begin{corollary}
If the functions in $\mathcal{F}$ are time-invariant,
Assumption \ref{ass:main:0} and \ref{ass:main:1} (2,3) hold and  $\sigma \in \mathcal{S}_{|\mathcal{F}|,D}$
is such that 
$\mathcal{G}_{\sigma(t)}$ is uniformly strongly connected,
then the consensus set $\mathcal{A}$ is globally quasi-uniformly attractive relative to $\mathcal{D}^*(\infty)$.
\end{corollary}

\begin{corollary}
If the functions in $\mathcal{F}$ are time-invariant,
Assumption \ref{ass:main:0} and \ref{ass:main:2} hold,
and $\sigma \in \mathcal{S}_{|\mathcal{F}|,D}$ it follows that the consensus set
$\mathcal{A}$ is globally uniformly asymptotically stable relative to  $\mathcal{D}^*(\infty)$
if and only if $\mathcal{G}_{\sigma(t)}$ is uniformly quasi-strongly connected
\end{corollary}

\section{Examples and interpretations}
In this section we provide some examples of 
systems on the form \eqref{chapter1:dynamics} for which the 
theorems are applicable.

\subsection{Non-convexity}
\begin{figure}[!htbp]
\centering
\includegraphics[width=0.6\columnwidth]{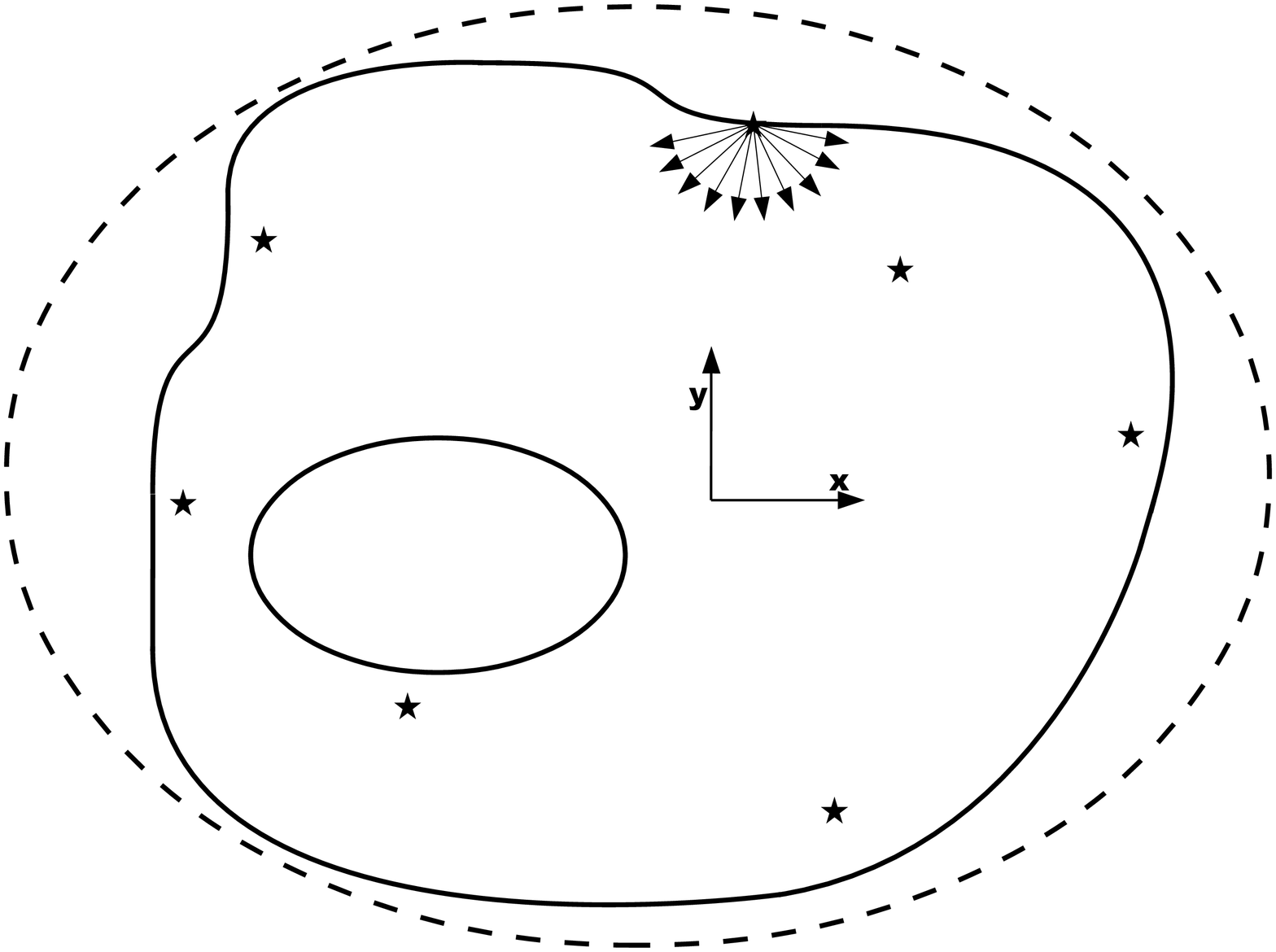}
\caption{Here we consider the case when $m =2$ and $n = 7$. The positions of the agents at a time $t$ are denoted by stars.
The solid curves is the set $\{y \in \mathbb{R}^m : V(y) = \alpha\}$. The dashed curve is the boundary of
$\mathcal{D}^*(\infty)$.}
\label{fig:two_regions}
\end{figure}

Suppose Assumption \ref{ass:main:0} is fulfilled and
there is a function $V$ such that Assumption \ref{ass:main:1}
is fulfilled for this $V$. In general the set $\{y \in \mathbb{R}^m : V(y) \leq \alpha\}$ does not need to be convex,
it depends on the function $V$.
This is illustrated in Figure \ref{fig:two_regions}, 
in which the two solid curves comprise the boundary of the set $\{y \in \mathbb{R}^m : V(y) \leq \alpha\}$
for some $\alpha > 0$. If all the agents are contained in this set at some time $t$ and there is an agent
$i$ on the boundary which has a neighbor $j$ such that $x_i \neq x_j$, then $x_i$ must move
away from the boundary into the interior of the set $\{y \in \mathbb{R}^m : V(y) \leq \alpha\}$. This is illustrated 
 in Figure \ref{fig:two_regions}, where the arrows indicate that the agent move into the interior of the
set $\{y \in \mathbb{R}^m : V(y) \leq \alpha\}$.

The dashed curve defines the boundary of the set $\mathcal{D}^{*}(\infty)$. Since the agents are contained 
in $\mathcal{D}^{*}(\infty)$ and $V$ fulfills Assumption \ref{ass:main:1}, provided $\mathcal{G}_{\sigma(t)}$ is uniformly
strongly connected, the system will reach consensus.

Another example where the theorems can be used is when the agents are contained in a geodesic convex and closed subset of a sphere. In 
this case we can choose $f_{W,m,m}(x_i, x_j)$ as the geodesic distance squared between 
$x_i$ and $x_j$. If $f_i(t,x)$
corresponds to a tangent vector that is 
inward-pointing~\cite{afsari2011riemannian} relative to the convex hull on the sphere (not to mix up with 
a convex hull in a Euclidean space) of the
positions of the neighbors of agent $i$ at time $t$
(provided it is nonempty otherwise $f_i(t,x) = 0$), then one can show that Assumption \ref{ass:main:2}
is fulfilled.

\subsection{Convexity}\label{sec:balls}
We continue with a less general case where
the decreasing functions are chosen as the Euclidean norm squared 
of the states and the relative states respectively. Under certain conditions, these choices
of functions can be used to show a well known convexity result
that, provided the right-hand side of each agent's dynamics as an element of the 
tangent space $T_{x_i}\mathbb{R}^{m}$ is inward-pointing~\cite{afsari2011riemannian}
relative to the convex hull of its neighbors, the system reaches consensus asymptotically~\cite{lin2007state,shi2009}.
We define the tangent cone to a convex
set $S \in \mathbb{R}^m$ at the point $y$ as
$$\mathcal{T}(y,S) = \bigg \{z \in \mathbb{R}^m: \liminf_{\lambda \rightarrow 0} \frac{\text{dist}(y + \lambda z, S)}{\lambda} = 0  \bigg \}.$$
This definition can be found in \cite{lin2007state}, and $\xi$ is inward-pointing relative to $S$, where $0 \neq \xi \in T_y\mathbb{R}^m$
($T_y\mathbb{R}^m$ is the tangent space of $\mathbb{R}^m$ at the point $y$),
if $\xi$ belongs to the relative interior of $\mathcal{T}(y,S)$. We use the term \emph{relative interior}, since the dimension of $S$ might be smaller
than $m$.  
Let us denote the convex hull for $\{x_i\}_{i = 1}^n$ by $\text{conv}(\{x_i\}_{i = 1}^n)$. 
Similarly, we can denote the convex hull for the positions of the neighbors of agent $i$ as
$\text{conv}(\{x_j\}_{j \in \mathcal{N}_i})$.

Suppose Assumption \ref{ass:main:0} is fulfilled.
We consider the case when $$V(x_i) = x_i^Tx_i \quad \text{and} \quad W(x_i, x_j) =  (x_j - x_i)^T(x_j- x_i),$$
where $V$ and $W$ generate the functions $f_{V,m}$ and $f_{W,m,m}$ respectively. 

\begin{figure}[!htbp]
\centering
\includegraphics[width=0.6\columnwidth]{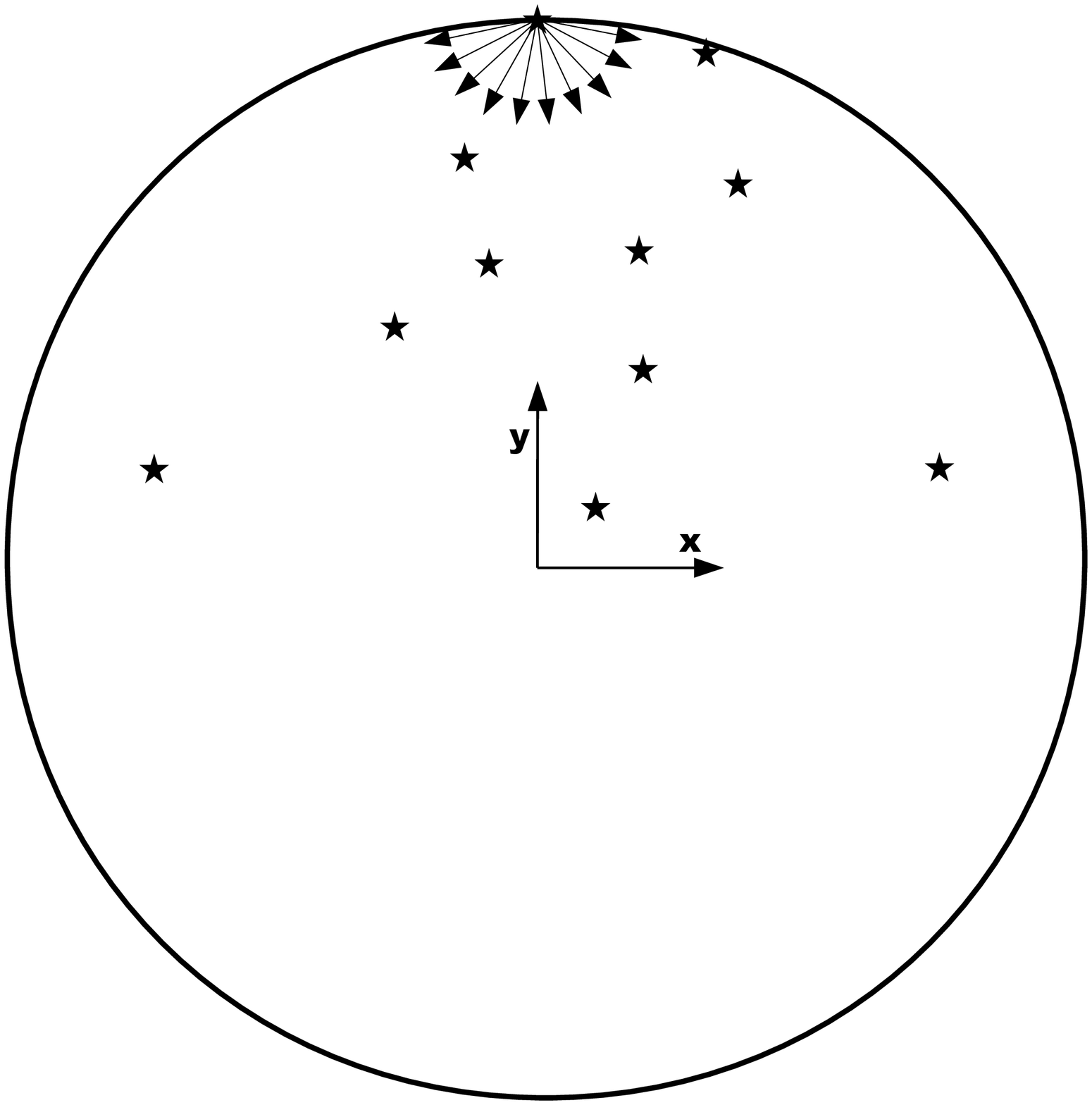}
\caption{In this case $m =2$. The positions of the agents at a time $t$ are denoted by stars. 
When at least one of the neighbors of an agent $i$ on the boundary of the ball 
$\bar{B}_{\max\limits_{k \in \mathcal{V}} \|x_k(t)\|,2}$  is located in the interior of the ball,
$f_i(t,x) \in T_{x_i}\mathbb{R}^2$ is inward-pointing (relative to the ball).}
\label{fig:sphere1}
\end{figure}

Suppose the functions in $\mathcal{F}$ are Lipschitz in $x$ on $\mathbb{R}^{mn}$, uniformly with respect to $t$,
and continuous in $t$.
Furthermore, suppose $V$ fulfills Assumption \ref{ass:main:1}, then in 
Theorem~\ref{chapter1:thm:1} we can choose $\widehat{\beta}_1(\|x_i\|) = \widehat{\beta}_2(\|x_i\|) = \|x_i\|^2$,
and obtain the result that any closed ball $\bar{B}_r$ in $\mathbb{R}^m$ is 
invariant and can be chosen as $\mathcal{D} = \mathcal{D}^*(\infty) = \bar{B}_r$, and the point $x = 0$ is uniformly stable. Thus, by Theorem \ref{chapter1:thm:3}
we obtain the result that if $\mathcal{G}_{\sigma(t)}$ is uniformly strongly connected, then
$\mathcal{A}$ is globally quasi-uniformly attractive relative to $\mathcal{D}^*(\infty)$.
Unless $x_i = x_j$ for all $ j \in\mathcal{N}_i$, for any agent $i$ that is furthest away from the origin, $f_i(t,x)$ 
as an element of the tangent space $T_{x_i}\mathbb{R}^m$ is inward-pointing on the boundary of 
the closed ball with radius equal to the norm of agent $i$.
This is illustrated in Figure 
\ref{fig:sphere1}.
An example of this situation is provided in Section~\ref{subsec:SO3}
in the application of reaching consensus for a system of rotation matrices. 

\begin{figure}[!htbp]
\centering
\includegraphics[width=0.6\columnwidth]{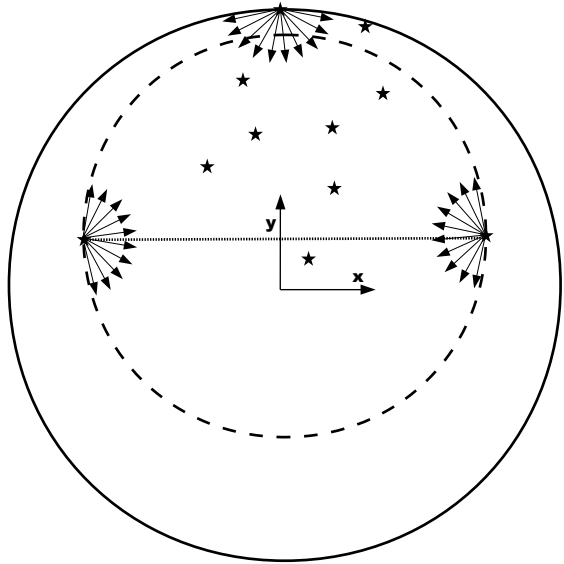}
\caption{In this case $m =2$. The positions of the agents at some time $t$ are denoted by stars.
The solid circle denotes the boundary of the ball with radius $\sqrt{f_{V,m}(x(t))}$ and the dashed circle
denotes the boundary of the ball with radius $\sqrt{f_{W,m,m}(x(t))}$.
The dashed line denotes the distance between the two agents that are furthest away from each other.}
\label{fig:sphere2}
\end{figure}

Suppose not only that $V$ fulfills Assumption \ref{ass:main:1}, but also that
 $W$ fulfills Assumption~\ref{ass:main:2}. In this case, any closed ball
in $\mathbb{R}^m$ is invariant and can be chosen as $\mathcal{D}^*(\infty)$, but also the largest Euclidean distance between any pair 
of agents is decreasing. This is illustrated in Figure~\ref{fig:sphere2}. 
Now Theorem~\ref{chapter1:thm:5} holds and $\mathcal{A}$ is globally uniformly asymptomatically stable relative to $\mathcal{D}^*(\infty)$ if 
and only if $\mathcal{G}_{\sigma(t)}$ is uniformly quasi-strongly connected.
For agent $i$, if $f_i(t,x)$ is inward-pointing relative to the convex hull of
its neighbors~\cite{lin2007state,shi2009}, 
then these conditions are fulfilled.

As a special case let 
$$f_i(t,x) = \sum_{j \in \mathcal{N}_i(t)}a_{ij}(t -\gamma_{\sigma}(t))(x_j - x_i),$$
where $\alpha_{ij}(t) > 0$ is continuous, positive and bounded for all $t$.
Let 
us construct the set of functions $\mathcal{F}$ in the following way. There are $2^{n^2}$ graphs. For each graph $\mathcal{G}_k$
we define a corresponding function 
$$\tilde{f}_k(x) = \left(\sum_{j \in \mathcal{N}_1}\alpha_{ij}(t)(x_j - x_1), \ldots, \sum_{j \in \mathcal{N}_n}\alpha_{ij}(t)(x_j - x_n) \right)^T$$
where $\mathcal{N}_i$ in this case is the neighborhood of agent $i$ in the graph $\mathcal{G}_k$. 
Now we let $$\mathcal{F} = \{\tilde{f}_k\}_{k=1}^{2^{n^2}},$$
and $\sigma \in \mathcal{S}_{|\mathcal{F}|,D,U}$. In the following examples,
if $\mathcal{F}$ is not explicitly defined, we assume that $\mathcal{F}$ is the set of functions that has been 
constructed in the way analogous to this construction, \emph{i.e.} all the possible right-hand sides.

 Now,
using the functions $$V(x_i) = x_i^Tx_i \quad \text{and} \quad W(x_i, x_j) =  (x_j - x_i)^T(x_j- x_i),$$
with the corresponding functions $f_{V,m}$ and $f_{W,m,m}$ respectively, one can show 
global uniform asymptotic consensus relative to $\mathcal{D}^*(\infty)$. 

\subsection{Nonlinear scaling}
Here we show how the theorems~\ref{chapter1:thm:3} and \ref{chapter1:thm:5} can be used
to assure consensus when the states and the relative states for pairs
of agents have been scaled with a nonlinear scale function.

In this context, let us define a nonlinear scale function as follows.
The function $g$ is strictly increasing on $[0,\eta)$ where $\eta > 0$ and the map
$$h: x_i \mapsto \frac{g(\|x_i\|)}{\: \: \|x_i\|}x_i$$
restricted to $B_{\eta,m}$ is a diffeomorphism between $B_{\eta,m}$ and $B_{\eta',m}$, where $\eta'~>~0$.

The interesting observation here regards the order of application of $h$. 
Suppose that
$$f_i(t,x) = \sum_{j \in \mathcal{N}_i(t)}a_{ij}(t -\gamma_{\sigma}(t))(x_j - x_i).$$
Within this context, if we define the following map
$${d}(x_i,x_j) = x_j - x_i,$$
we can write the function $f_i$ as follows
$$f_i(t,x) = \sum_{j \in \mathcal{N}_i(t)}a_{ij}(t-\gamma_{\sigma}(t)){d}(x_i,x_j),$$
and we know that $f_i$, as an element of the tangent space $T_{x_i}\mathbb{R}^m$, is inward-pointing relative to 
the convex hull of the neighbors
of agent $i$. Consequently, on $B_{\eta,m}$, we can use Theorem~\ref{chapter1:thm:3} together with Theorem~\ref{chapter1:thm:5} in order to show
consensus when the graph $\mathcal{G}(t)$ is uniformly quasi-strongly connected. 
Now, for each pair of agents, if we modify $f_i$ into the following form
$$f_i'(t,x) = \sum_{j \in \mathcal{N}_i(t)}a_{ij}(t - \gamma_{\sigma}(t))h({d}(x_i,x_j)), $$
this new function still fulfills the same convexity assumption.
However, if we reverse the order of application of the functions $h$ and ${d}$ we get the following modified version
of $f_i$
$$f_i''(t,x) = \sum_{j \in \mathcal{N}_i(t)}a_{ij}(t-\gamma_{\sigma}(t)){d}(h(x_i),h(x_j)),$$
and in this case it is not necessarily true that $f_i''(t,x)$ as an element of $T_{x_i}\mathbb{R}^m$
is inward-pointing relative to the convex hull of the neighbors of agent $i$. 
However, consensus can be guaranteed on $B_{\eta,m}$ by Theorem~\ref{chapter1:thm:3} when the graph $\mathcal{G}(t)$
is uniformly strongly connected by using the function $V(x_i) =x_i^Tx_i$ in Theorem~\ref{chapter1:thm:3}.

\subsection{Avoiding discontinuities}
Suppose that $\mathcal{F}$ contains only time-invariant functions, $\sigma \in \mathcal{S}_{|\mathcal{F}|, D,U}$ and
Assumption~\ref{ass:main:0} holds.
We show how it is possible to modify the system defined by $\mathcal{F}$ and $\sigma$ into a system where the right-hand
side is no longer discontinuous in $t$. Close to each switching time we can modify the system so that there is a continuous in time transition between
the two time-invariant functions that are being switched between. For the modified system where there are no longer any discontinuities in $t$,
Assumption~\ref{ass:main:0} still holds and if there is a $V$ such that Assumption~\ref{ass:main:1}
holds for this $V$ for the discontinuous system (or a $W$ such that Assumption~\ref{ass:main:2} holds for this $W$ for the 
discontinuous system), then Assumption~\ref{ass:main:1} holds for $V$ (or Assumption~\ref{ass:main:2} holds for $W$) for the modified
continuous system. 

We start by extending $\mathcal{F}$ with time 
varying functions to a finite set of functions $\mathcal{F}'$ (Lipschitz in $x$ on $\mathcal{D}$, uniformly with respect to $t$),
where $\mathcal{F}'$ contains functions that serve as continuous in time transitions between functions in $\mathcal{F}$.
For $\sigma \in \mathcal{S}_{|\mathcal{F}'|,D,U}$ we
create a $\sigma' \in \mathcal{S}_{|\mathcal{F}'|,D,U}$ in the following way. Let $\tau_D^{\sigma'} < \tau_D^{\sigma}$.
At each switching time $\tau_k$ of $\sigma$, we squeeze in an extra interval of 
length $\tau_D^{\sigma'}$
during which the neighbor set $\mathcal{N}^{\sigma'}_i$ of each agent $i$ is equal to
$\mathcal{N}^{\sigma}_i(\tau_{k-1}) \cup \mathcal{N}^{\sigma}_i(\tau_{k})$. 
These added time intervals can be seen as transition periods,
during which there is a continuous in time transition between two functions in $\mathcal{F}$.

We extend $\mathcal{F}$ to $\mathcal{F}'$ in the following way. 
First we define a continuous function 
$$\alpha : (-\infty, \infty) \rightarrow [0,1],$$
such that $\alpha(0) = 1$ and $\alpha(\tau_D^{\sigma'}) = 0$.
Secondly, for each pair of functions $(\tilde{f}_i,\tilde{f}_j)$ where $\tilde{f}_i$ and
$\tilde{f}_j$ belong to $\mathcal{F}$, we define a function 
$$\tilde{f}_{(i,j)}(t,x) = \alpha(t)\tilde{f}_i(x) + (1-\alpha(t))\tilde{f}_j(x).$$
The set of functions $\mathcal{F}'$ is the set of all functions $\tilde{f}_i$ and $\tilde{f}_{(i,j)}$.
At each switching time of the original system, between the right-hand side $\tilde{f}_i$
and $\tilde{f}_j$, 
we now squeeze in the function $\tilde{f}_{(i,j)}$ during a time period of length $\tau_D^{\sigma'}$ in the
new system. 
Note that we can make $\tau_D^{\sigma'}$ much smaller than $\tau_D^{\sigma}$.

If all functions in $\mathcal{F}$ are time-invariant $\mathcal{C}^1$ functions in $x$, and 
we want the new continuous right-hand side to be $\mathcal{C}^1$ in $t$
when $x$ is regarded as a function of $t$, we 
impose the additional requirement that $\dot{\alpha}(0) = 0$ and
$\dot{\alpha}(\tau_D^{\sigma'}) = 0$. A function fulfilling these requirements
is 
$$\alpha(t) =  \frac{1}{2} + \frac{1}{2}\cos \left(\frac{t\pi}{\tau_D^{\sigma'}} \right) .$$

We now proceed with some other application oriented examples.

\subsection{Consensus on $SO(3)$ using the Axis-Angle Representation}\label{subsec:SO3}
Here we have a system of $n$ rotation matrices in $SO(3)$ (controlled on a kinematic level) shall asymptotically 
reach consensus in the rotation matrices. 
For a rotation matrix $R_i$ there is a corresponding vector $x_i$,
referred to as the Axis-Angle Representation of $R_i$. 
Locally around the identity matrix, in terms of kinematics we have that
$$\dot{R}_i = R_i\widehat{\omega}_i \quad \quad \text{or} \quad \quad \dot{x}_i = {L}_{x_i}\omega_i,$$
where 
\begin{equation*}
{L}_{{x}_i} = I_3 +
\frac{\widehat{{x}}_i}{2}+ \frac{1}{\|{x}_i\|^2}\bigg (1 -
\frac{\text{sinc}(\|{x}_i\|)}{\text{sinc}^2(\frac{\|{x}_i\|}{2})}\bigg
)\widehat{x}_i^2,
\end{equation*}
and $\widehat{\omega}_i$, $\widehat{x}_i$ are the skew-symmetric matrices generated by $\omega_i, x_i \in \mathbb{R}^3$ respectively,
 and we require that
 $x_i(t_0) \in B_{\pi,3}$ for all $ i$.
Now we consider the case when $$\omega_i = \sum_{j \in \mathcal{N}_i(t)}\alpha_{ij}(t - \gamma_{\sigma}(t))(x_j - x_i),$$
where the continuous function $\alpha_{ij}(t)$ is positive and bounded, and $\sigma \in \mathcal{S}_{|\mathcal{F}|,D,U}$. 
 The symmetric part of the matrix ${L}_{x_i}$ is positive definite 
on $B_{\pi,3}$, and the system is at
an equilibrium if and only if $x = (x_1, \ldots, x_n)^T \in \mathcal{A}$. 

Let $V(x_i) = x_i^Tx_i$. By observing that $x_i^T{L}_{x_i} = x_i^T$,
it is easy to show that Assumption \ref{ass:main:1} holds for $V$. We can apply
Theorem \ref{chapter1:thm:1} with $\widehat{\beta}_1(\|x_i\|) = \widehat{\beta}_2(\|x_i\|) = \|x_i\|^2$, and show that
any ball $\bar{B}_{r,3}$ is invariant for $r < \pi$ and may serve as $\mathcal{D} = \mathcal{D}^*(\infty)$. Also, by Theorem \ref{chapter1:thm:3}, if the graph $\mathcal{G}_{\sigma(t)}$
is uniformly strongly connected, then $\mathcal{A}$ is globally quasi-uniformly 
attractive.

\subsection{Consensus on $SO(3)$ for networks of cameras using the epipoles}
This example is based on the work in \cite{montijanoepipolar,montijano2011multi},
where a more detailed description can be obtained. Undefined terminology that is used in this
example can be found in any standard text book on computer vision such as~\cite{ma2004invitation}.
This example also regards consensus for rotation matrices, but the setting is a
bit different and the rotations are restricted to be only around one common axis.
 We consider a system of $n$ robots positioned 
in the two-dimensional plane. Each robot is equipped with a camera
and is at each time observing a subset of the other robots. Since
the rotational axes are fixed and equal, we only need the scalar $\theta_i$
in order to represent the rotation of each agent $i$,
where $\theta_i$ is the angle of rotation.
In the context of this example, instead of letting $\theta_i \in [0, \pi)$,
we let $\theta_i \in (-\pi, \pi)$.
We assume that all the cameras have the same intrinsic parameters.

The robots are not moving and are only rotating.
The position of each robot $i$ in the world coordinate
frame is given by $x_i \in \mathbb{R}^2$. 
The position of agent $j$ in the body frame of agent $i$ is given by
$$x_{ij}(\theta_i) = R(\theta_i)(x_j - x_i),$$
where $$ R(\theta_i) = \begin{bmatrix}
\cos(\theta_i) & -\sin(\theta_i) \\
\sin(\theta_i) & \cos(\theta_i)
\end{bmatrix}.$$
Let $$\psi_{ij}(\theta_i) = \text{arctan}\left(\frac{x_{ijx}}{x_{ijy}}\right),$$
where $x_{ijx}$ and  $x_{ijy}$ are the two components of $x_{ij}$.

Now, instead of measuring the rotation directly,
using stereo vision one retrieves the \emph{epipoles}
as certain nullspace vectors of 
the so called fundamental matrix. The fundamental matrix defines
the (epipolar) geometric relationship between two images~\cite{ma2004invitation}, and should not be
mixed up with the fundamental matrix in the solution of a linear
time-invariant dynamical system. We will only consider 
the $x$-component (the first component) of these two-dimensional epipole vectors,
which are defined as 
$$e_{ij} = \alpha\tan(\psi_{ij}), \quad e_{ji} = \alpha\tan(\psi_{ij} - \theta_{ij}),$$
where $\theta_{ij} = \theta_j - \theta_i$ and  $\alpha = 1$ if the cameras are calibrated,  \emph{i.e.}, the focal length is known (we assume that
the position of the principal point is known in the image plane), otherwise $\alpha > 0$ is unknown. 

Let us define $$\omega_{ij} = \arctan\left(\frac{e_{ij}}{\beta}\right) - \arctan\left(\frac{e_{ji}}{\beta}\right)$$
where $\beta > 0$ is a constant to choose. 

We define $\theta(t) = (\theta_1(t), \ldots, \theta_n(t))^T$ and the region
$$\mathcal{D} = \{\theta :-\theta_M \leq \theta_i \leq \theta_M \text { for } i = 1, \ldots, n\},$$ where
$0 < \theta_M \ll \pi /2$. The set $\mathcal{D}$ could be seen as being a function of $\theta_M$. Furthermore, we assume
${x_{ijx}(0)}/{x_{ijy}(0)} = 1$ for all $i,j$, in which case the robots or the cameras are standing on a line and are
oriented in the same direction that forms an angle of $\pi/4$ to 
the direction of the line. This means that $\psi_{ij} \in \{-\pi/4,  3 \pi /4 \}$ for all $i,j$.

Let us choose the dynamics for the system as
\begin{align*}
\dot{\theta}_1 = &\sum_{j \in \mathcal{N}_1(t)}\alpha_{1j}(t - \gamma_{\sigma}(t))\omega_{1j}, \\
& \vdots \\
\dot{\theta}_n =& \sum_{j \in \mathcal{N}_n(t)}\alpha_{nj}(t -  \gamma_{\sigma}(t))\omega_{nj}.
\end{align*}
We assume that $\alpha_{ij}(t)$ is continuous, positive and bounded, and $\sigma \in \mathcal{S}_{|\mathcal{F}|,D,U}$. 
Provided $\theta_M$ is sufficiently small, on $\mathcal{D}$ it can be shown that $\omega_{ij}$ is Lipschitz for all $(i,j) \in \mathcal{V} \times \mathcal{V}$.
It is obvious that Assumption~\ref{ass:main:0} holds.
We choose $\theta_M$ small enough so that $|\omega_{ij}| < \pi/2$ on $\mathcal{D}$. 
According to \cite{montijanoepipolar}, it is true that 
\begin{equation}\label{eq:nisse1}
\theta_{ij}(t) \neq 0 \Longrightarrow \theta_{ij}(t)\omega_{ij}(t) > 0.
\end{equation}

Let us now consider the function $V(\theta_i) = \theta_i^2$, 
where 
$$\frac{d}{dt}{V}(\theta_i) = 2\theta_i\sum_{j \in \mathcal{N}_i(t)}\alpha_{ij}(t)\omega_{ij}.$$
Suppose $i \in \mathcal{I}_V(t,t)$,
and $\theta_j(t) = \theta_i(t)$ for all $j \in \mathcal{N}_i(t)$, then it follows that 
$\dot{V}(\theta_i(t)) = 0$. Now, consider the 
situation where $i \in \mathcal{I}_V(t,t)$ and there is 
at least one $j$ such that 
 $\theta_j(t) \neq \theta_i(t)$ when $j \in \mathcal{N}_i(t)$. Since $i \in \mathcal{I}_V(t,t)$,
if $\theta_i \neq \theta_j$, using \eqref{eq:nisse1} we get that
$$\theta_i\omega_{ij} < 0.$$
Hence, Assumption \ref{ass:main:1} also holds. 

In Theorem~\ref{chapter1:thm:1} we can now choose $\widehat{\beta}_1(|\theta_i|) = 
\widehat{\beta}_2(|\theta_i|) = |\theta_i|^2$ and reach the conclusion that
$\mathcal{D}$ is positively invariant and $\mathcal{D}^*(\infty) = \mathcal{D}$. The point $0$ is uniformly stable. Furthermore,
according to Theorem~\ref{chapter1:thm:3}, $\mathcal{A}$ is globally quasi-uniformly attractive relative
to $\mathcal{D}^*(\infty)$
if  $\mathcal{G}_{\sigma(t)}$ is uniformly strongly connected. But we can actually weaken
the assumptions on the graph $\mathcal{G}_{\sigma(t)}$.

Let us consider the function $W(\theta_i, \theta_j) = (\theta_j - \theta_i)^2$,
where 
\begin{align*}
& \frac{d}{dt}{W}(\theta_i, \theta_j)  \\
& = 2(\theta_j - \theta_i)\left(\sum_{k \in \mathcal{N}_j(t)}\alpha_{jk}(t - \gamma_{\sigma}(t))\omega_{jk} - \sum_{l \in \mathcal{N}_l(t)}\alpha_{il}(t- \gamma_{\sigma}(t))\omega_{il}\right).
\end{align*}
 If $(i,j) \in \mathcal{J}_V(t,t)$, we can without loss of generality assume that $\theta_j \geq \theta_k$
and that $\theta_i \leq  \theta_k$ for all $k \in \mathcal{V}$. This implies that $\text{sign}(\theta_{ij}) =  \text{sign}(\theta_{kj}) = \text{sign}(\theta_{il})=1$ for all $k,l \in \mathcal{V}$, so from \eqref{eq:nisse1} we get that 
$\text{sign}(\theta_{ij})\text{sign}(\omega_{jk}) = -1$ and $\text{sign}(\theta_{ij})\text{sign}(\omega_{il}) = 1$.
Thus Assumption  \ref{ass:main:2} holds for $f_{W,m,m}$
and Theorem \ref{chapter1:thm:5} can be used. Thus, when $x(t) \in \mathcal{D}^*(\infty)$ it follows that
 $\mathcal{A}$ is globally uniformly asymptotically stable relative to $\mathcal{D}^*(\infty)$
if and only if $\mathcal{G}_{\sigma(t)}$ is uniformly quasi-strongly connected.

\subsection{Stabilization}
Let us now, as a special case of the consensus problem, consider the stabilization problem, where we use
our consensus results in order to provide known conditions for when
$\{0\}$ is asymptotically stable for a system
\begin{align}\label{eq:stabilization_dynamics}
\dot{y} = g(t,y),
\end{align}
where $y \in \mathbb{R}^m$.  

We show that 
this problem is a special case of a consensus problem with two agents in $\mathbb{R}^m$,
so that we can use Theorem~\ref{chapter1:thm:5} in order to show that $\{0\}$
is globally uniformly asymptotically stable relative to some compact invariant set in $\mathbb{R}^m$. 

\begin{proposition}\label{cor:75}
Suppose there is an invariant compact set $\mathcal{D}' \subset \mathbb{R}^m$ containing 
the point $0$
and a finite set $\mathcal{F}' = \{\tilde{f}_1', \ldots, \tilde{f}_{|\mathcal{F}'|}' \}$ of functions that are piecewise 
continuous in $t$ and Lipschitz in $y$ on $\mathcal{D}'$, uniformly with respect to $t$. For each function $\tilde{f}_k'$ it holds that
$\tilde{f}_k'(t,0) = 0$ for all $k$ and all $t$. Furthermore, 
$\sigma \in \mathcal{S}_{|\mathcal{F}'|,D,U}$ and the right-hand side of \eqref{eq:stabilization_dynamics} is
$$g(t,y) = \tilde{f}_{\sigma(t)}'(t - \gamma_{\sigma}(t),y).$$

If there is a positive definite function
$V(y)$, which is continuously differentiable on an open set containing $\mathcal{D}'$ such that $$\nabla V(y)\tilde{f}_i'(t,y) < 0$$ for all
$i \in \{1, \ldots, |\mathcal{F}'|\}$, all $t$ and nonzero $y$ in $\mathcal{D}'$, then 
$\{0\}$ is globally uniformly asymptotically stable relative to $\mathcal{D}'$.
\end{proposition}

\emph{\quad Proof}: 
The set $\mathcal{D}'$ is assumed to be invariant for any choices of switching signal functions
in $\mathcal{S}_{|\mathcal{F}'|,D}$.
Let us define a system of two agents, agent 1 and agent 2. 
Based on the set $\mathcal{F}'$ we create a new set $\mathcal{F}''$ of functions
with range $\mathbb{R}^{2m}$
in the following way
$$\mathcal{F}'' = \{(\tilde{f}'_1(t,y_2 - y_1), 0)^T, \ldots, (\tilde{f}'_{|\mathcal{F}'|}(t,y_2 - y_1), 0)^T\}.$$
Now, for all $t \geq 0$ and for all $\sigma \in \mathcal{S}_{|\mathcal{F}''|,D}$ we define 
\begin{align*}
\mathcal{N}_1^{\sigma}(t)  = \{1,2\} \quad \text{ and } \quad
\mathcal{N}_2^{\sigma}(t)  =  \{2\}.
\end{align*}
The system dynamics for this extended system is given as
\begin{align*}
\dot{y}_1 & = \tilde{f}'_1(t - \gamma_{\sigma}(t),y_2 - y_1), \\
\dot{y}_2 & = 0. \\
\end{align*}
This system fulfills Assumption~\ref{ass:main:0} and we define a function $W$ as
$$W(y_1,y_2) = V(y_2-y_1).$$
The function $W$ fulfills Assumption~\ref{ass:main:2}. Now, if the initial positions of 
$y_1(t)$ and $y_2(t)$ are $y_1^0 \in \mathcal{D}'$ and $y_2^0 = 0 \in \mathcal{D}'$ respectively, we see that the dynamics
for the extended system is
equivalent to the original system \eqref{chapter1:dynamics}. For the extended system,
the set $\mathcal{D}' \times \{0\} \subset ((\mathcal{D}')^2)^*(\infty)$.
Since $\mathcal{G}_{\sigma(t)}$
is uniformly quasi-strongly connected, $\mathcal{A}$ is globally uniformly asymptotically 
stable relative to $((\mathcal{D}')^2)^*(\infty)$. Since $y_2(t) = 0$ for all $t$, we see that
 the state will converge to the point $(0,0)^T \in \mathbb{R}^{2m}$ in the extended system.
\hfill $\blacksquare$ \vspace{2mm}

\section{Proofs}\label{sec:proofs}
In this section we provide the proofs.
Theorem \ref{chapter1:thm:1} is proven directly, whereas for the two
other theorems, in order
to make the proofs more comprehensible, we first introduce some lemmas,
used as building blocks for the final proof.

\vspace{3mm}
\quad \emph{Proof of Lemma~\ref{lem:a1}}:   
We can construct the $\sigma'$ as follows.
Let us first choose $\tau_D^{\sigma'} = \tau_D^{\sigma}$ and $\tau_U^{\sigma'} \geq 2\tau_D^{\sigma}$. For any $k$ such that 
$\tau_{k+1} - \tau_k > \tau_U^{\sigma'}$, we split $[\tau_{k}, \tau_{k+1})$ into a partition
of smaller half-open intervals each with equal length smaller than $\tau_U^{\sigma'}$ but larger than $\tau_D^{\sigma'}$. On these
half-open intervals $\sigma'(t) = \sigma(t)$. For all $k$ such that $\tau_{k+1} - \tau_{k} \leq \tau_{U}^{\sigma'}$
we let $\sigma'(t) = \sigma(t)$ for $t \in [\tau_k, \tau_{k+1})$.
\hfill $\blacksquare$

\vspace{3mm}
\quad \emph{Proof of Lemma~\ref{lem:a2}}:   
Let $\tau_U = \tau_U^{\sigma}$ and $\tau_D = \tau_D^{\sigma}$.
The function $\sigma'$ is constructed in a way similar to the procedure in the proof of Lemma~\ref{lem:a1},
but here the number of half-open intervals that $[\tau_{k}, \tau_{k+1})$ is split into is bounded from 
above by $	\lfloor \tau_U/\tau_D \rfloor$. 

We define the partition of intervals as follows
\begin{align*}
[\tau_{k}, \tau_{k+1})= &\left(\bigcup_ {i= 1}^{\lfloor (\tau_{k+1} - \tau_k)/\tau_D \rfloor -1}[\tau_{k} + (i-1)\tau_D, \tau_{k} + i\tau_D) \right)\\
&\:\: \: \quad \cup [\tau_{k} + (\lfloor (\tau_{k+1} - \tau_k)/\tau_D \rfloor -1)\tau_D,\tau_{k+1}).
\end{align*}

We define $\mathcal{F}'$ as follows
\begin{align*}
\mathcal{F}'  =  \{& \tilde{f}_1' = \tilde{f}_1(t,x), \\
&  \tilde{f}_2' = \tilde{f}_1(t + \tau_D,x), \ldots, \\
&\tilde{f}_{\lfloor \tau_U/\tau_D \rfloor}' = \tilde{f}_1(t + (\lfloor \tau_U/\tau_D \rfloor -1)\tau_D,x), \ldots, \\
& \tilde{f}_{\lfloor \tau_U/\tau_D \rfloor (N-1) + 1}' =\tilde{f}_N(t + \tau_D,x), \ldots, \\
& \tilde{f}_{\lfloor \tau_U/\tau_D \rfloor N}' = \tilde{f}_N(t +  (\lfloor \tau_U/\tau_D \rfloor -1)\tau_D,x)\},
\end{align*}
where $N = |\mathcal{F}|$.
The set $\mathcal{F}'$ is constructed by creating $\lfloor \tau_U/\tau_D \rfloor-1$
number of new time-shifted functions from each function $\tilde{f}_i \in \mathcal{F}$. 

Now $\sigma'$ is constructed by choosing a function in $\mathcal{F}'$ on each half-open interval
in each partition
so that $$\tilde{f}_{\sigma(t)}(t,x) = \tilde{f}_{\sigma'(t)}'(t,x)$$ for all $t$ and $x \in \mathcal{D}$.
\hfill $\blacksquare$

\vspace{3mm}
\quad \emph{Proof of Lemma~\ref{lem:dini}}:   
We only prove the first statement for $f_{{V},m}$, the procedure in order to prove 
the second statement for $f_{{V},m,m}$ is similar and hence omitted.

Since $V$ is Lipschitz in $x$ on $\mathcal{D}$ it follows that $f_{V,m}$ is Lipschitz in $x$
on $\mathcal{D}$. Since $f_{V,m}$ is Lipschitz in $x$, it follows that
$$D^+(f_{V,m}(x(t)))= D^+_{f_{V,m}}(f_{V,m}(x^*)),$$
where 
$$D^+_{f_{V,m}}(f_{V,m}(x^*)) = \lim_{\epsilon \downarrow 0}\sup\frac{ f_{f_{V,m}}(t + \epsilon, x_0 + \epsilon f_{V,m}(t,x^*))}{\epsilon}$$
and $x^* = x(t)$. This result can be obtained from Chapter 1 in \cite{yoshizawa1966stability}. In \cite{rouche1977stability}
it is formulated as a Theorem (Theorem~4.1 in Appendix I).

The next step is to prove that
$$D^+_{f_{V,m}}(f_{V,m}(t,x^*)) = \max_{i \in \mathcal{I}_{V}(t,t)}\frac{d}{dt}{V}(x_i(t)).$$
This result can for example be obtained from 
Theorem~2.1. in \cite{clarke1975generalized}.
\hfill $\blacksquare$

\vspace{3mm}
\quad \emph{Proof of Lemma \ref{lem:79}}:   
Since $\mathcal{D}$ is compact, we only need to verify that 
$\mathcal{D}^*(\tilde{t})$ is closed in order to show that $\mathcal{D}^*(\tilde{t})$
is compact.
Suppose there is $x_0 \notin \mathcal{D}^{*}(\tilde{t})$,
such that there is a sequence $\{x_0^i\}_{i= 1}^{\infty}$ that converges to 
$x_0$, where each element in the sequence is in $\mathcal{D}^{*}(\tilde{t})$. We would like to obtain a contradiction by showing that 
the solution $x^{\sigma}(t,t_0,x_0)$
does exist in $\mathcal{D}$ on the interval $[t_0, t_0 + \tilde{t})$ for any $t_0$, and $\sigma \in \mathcal{S}_{|\mathcal{F}|,D}$.

By using the fact that $\mathcal{D}$ is compact and that the right-right side of \eqref{chapter1:dynamics} is uniformly Lipschitz in $x$
on $\mathcal{D}$ and piecewise continuous in $t$, we can use the Continuous Dependency Theorem 
of initial conditions in order to guarantee that
$\{x^{\sigma}(t,t_0,x_0^i)\}_{i=1}^{\infty}$ is a Cauchy sequence for arbitrary $t \in [t_0, t_0 +\tilde{t})$.
Now we know, since $\mathcal{D}$ is compact, that
$$x^*(t) = \lim_{i \rightarrow \infty}x^{\sigma}(t,t_0,x_0^i)$$
exists and $x^*(t) \in \mathcal{D}$.
We want to prove that $x^*(t)$ is the solution for \eqref{chapter1:dynamics} on $[t_0, t_0 + \tilde{t})$ 
for the given $\sigma$, $t_0$ and $x_0$.
\begin{align*}
x^*(t) & = \lim_{i \rightarrow \infty}x^{\sigma}(t,t_0,x_0^i) \\
& =  \lim_{i \rightarrow \infty}\int_{t_0}^tf(s,x^{\sigma}(s,t_0,x_0^i))ds \\
& = \int_{t_0}^t\lim_{i \rightarrow \infty}f(s,x^{\sigma}(s,t_0,x_0^i))ds \\
& = \int_{t_0}^tf(s,x^*(s)).
\end{align*}
Hence, $x^*(t)$ is contained $\mathcal{D}$ for all $t$, but since $\sigma$ and $t_0$ were
arbitrary, it follows that $x_0 \in \mathcal{D}^{*}(\tilde{t})$
which is a contradiction.

Now we prove the statement that  $\mathcal{D}^{*}(\infty)$ is invariant.
Suppose $x_0 \in \mathcal{D}^{*}(\infty)$ is arbitrary and let 
$$y = x^{\sigma_1}(t_1,t_0,x_0)$$
for $\sigma_1 \in \mathcal{S}_{|\mathcal{F}|,D}$ and $t_1 \geq t_0$. 
Consider
$x^{\sigma_2}(t,t_1',y)$ for some arbitrary $\sigma_2 \in \mathcal{S}_{|\mathcal{F}|,D}$
and $t_1'$. We need to show that $x^{\sigma_2}(t,t_1',y)$ is contained in $\mathcal{D}$
for all $t \geq t_1'$. 

We define 
$$\sigma(t) = \begin{cases}
\sigma_1(t-(t_1' - t_1)) & \text{if } t < t_1', \\
\sigma_2(t) & \text{if } t \geq t_1'.
\end{cases}
$$
which is contained in $\mathcal{S}_{|\mathcal{F}|,D}$. Thus
$$x^{\sigma_2}(t,t_1',y) = x^{\sigma}(t,t_0 + (t_1' - t_1),x_0)$$
which is contained in $\mathcal{D}$ for all $t \geq t_0$ since $x_0 \in \mathcal{D}^*(\infty)$. Thus $y \in \mathcal{D}^*(\infty)$.
\hfill $\blacksquare$

\vspace{3mm}
\quad \emph{Proof of Theorem \ref{chapter1:thm:1}}:   
Since the origin is an interior point of $\mathcal{D}$, there is a ball 
${B}_{\epsilon,m}$
such that $({B}_{\epsilon,m})^n \subset \mathcal{D}$ and $\epsilon > 0$. 
Suppose $x_0 \in ({B}_{\epsilon,m})^n$, then there is a closed ball 
$$\bar{B}_{\epsilon',mn}(x_0) \subset ({B}_{\epsilon,m})^n$$
with $\epsilon' > 0$.
Now according to Theorem 3.1. in \cite{khalil2002nonlinear}, there 
is a $\delta' > 0$ such that the system has a unique solution ${x}(t,t_0,x_0)$ on $[t_0, t_0 + \delta']$. 
We choose $[t_0, t_0 + {T'})$ as the maximal half-open interval of existence of the unique solution.
We know there are class $\mathcal{K}$ functions $\beta_1$ and $\beta_2$ such that
$$\beta_1(\|y\|) \leq V(y) \leq \beta_2(\|y\|)$$ 
for $y \in \mathbb{R}^m$. 

Now, using property (2) of Assumption \ref{ass:main:1} we get from the 
Comparison Lemma (Lemma 3.4 in~\cite{khalil2002nonlinear}),
that $$f_{V,m}({x}(t)) \leq f_{V,m}({x}_0)$$ for $t \in [t_0, t_0 + {T}').$
Now let $\delta = \beta_2^{-1}(\beta_1(\epsilon))$. 
We suppose that $x_0$ was chosen such that 
$$x_i(t_0) \in \bar{B}_{\delta,m} \subset \bar{B}_{\epsilon,m} \text{ for all } i.$$ It follows that for $t \in [t_0, t_0 + T')$,
\begin{align*}
&\max_{i \in \{1, \ldots, n\}}\|x_i(t)\|  = \beta_1^{-1}(\beta_1(\max_{i \in \{1, \ldots, n\}}\|x_i(t)\|)) \\
&= \beta_1^{-1}(\max_{i \in \{1, \ldots, n\}}\beta_1(\|x_i(t)\|))  \leq \beta_1^{-1}(f_{V,m}(x(t)))\\
 & \leq  \beta_1^{-1}(f_{V,m}(x(t_0))) \leq  \beta_1^{-1}(\max_{i \in \{1, \ldots, n\}}\beta_2(\|x_i(t_0)\|)) \\
&  \leq  \beta_1^{-1}(\beta_2(\max_{i \in \{1, \ldots, n\}}(\|x_i(t_0)\|)))  \leq  \beta_1^{-1}(\beta_2(\delta)) = \epsilon.
\end{align*}
Now it follows 
by using Theorem 3.3 in~\cite{khalil2002nonlinear}, that the solution will stay in $(\bar{B}_{\epsilon,m})^n$
for arbitrary times larger than $t_0$, \emph{i.e.}, $T' = \infty$.
\hfill $\blacksquare$ \vspace{2mm}

In the following lemma we use the positive limit set $L^+(x_0, t_0)$ of 
the solution \newline
$x(t,t_0, x_0)$ when $x_0 \in \mathcal{D}^*(\infty)$ (we assume that $\sigma \in \mathcal{S}_{|\mathcal{F}|,D}$ is fixed here).
 This limit set exists and is compact, and $x(t)$ approaches
 it as the time goes to infinity, however it is not guaranteed to be invariant which is 
the case for an autonomous system. Now, in the case that $x_0 \in \mathcal{D}^*(\infty)$,
the set $L^+(x_0, t_0)$ is contained in $\mathcal{D}^*(\infty)$, so any alternative solution of \eqref{chapter1:dynamics} 
that starts in $L^+(x_0, t_0)$ will remain in $\mathcal{D}^*(\infty)$.

\begin{lemma}\label{lem:11}
Suppose that $x_0 \in \mathcal{A}^c  \cap \mathcal{D}^*(\infty)$ and that
 Assumption \ref{ass:main:1} (2) holds.
Suppose that
there is a non-negative function $\beta(y, \tilde{t})$ 
that is increasing in $\tilde{t}$ for $y \in \mathcal{A}^c \cap \mathcal{D}^*(\infty)$. Furthermore, suppose that
for $y \in \mathcal{A}^c \cap \mathcal{D}^*(\infty)$,
there is $\tilde{t}'(y) >0$ such that for $\tilde{t} \geq \tilde{t}'(y)$ it holds that $\beta(y, \tilde{t}) > 0$.

If
$$f_{V,m}(x(t_0 + \tilde{t}, t_0, x_0)) - f_{V,m}(x_0) \leq -\beta(x_0, \tilde{t}),$$
then $x(t) \rightarrow \mathcal{A}$ as $t \rightarrow \infty$ for all $t_0$.

Furthermore, if $\beta$ is lower semi-continuous in $y$, and $\tilde{t}'$ is 
independent of $y$, then $\mathcal{A}$
is globally quasi-uniformly attractive relative to $\mathcal{D}^*(\infty)$.
\end{lemma}

\vspace{3mm}
\quad \emph{Proof}:   
Let us consider an arbitrary $x_0 \in \mathcal{A}^c  \cap \mathcal{D}^*(\infty)$ and $t_0$ for which the 
solution $x(t,t_0, x_0)$ generates the limit set $L^+(x_0, t_0) \subset \mathcal{D}^*(\infty)$.
From the fact that $f_{V,m}(x(t))$ is continuous in $t$, the fact that $f_{V,m}(x(t))$ is decreasing and the fact that $x(t)$ is contained in the compact
set $\mathcal{D}^*(\infty)$, it follows that $f_{V,m}(x(t,t_0,x_0))$
converges to a lower bound $\alpha(x_0,t_0) \geq 0$ as $t\rightarrow \infty$. 
Suppose $L^+(x_0, t_0) \not\subset \mathcal{A}$. We want to prove the lemma by 
showing that this assumption leads to a contradiction. Let $t_1 \geq t_0$ be arbitrary 
and $y_1$ be an arbitrary point in $L^+(x_0, t_0) \cap \mathcal{A}^c$.
Since $y_1 \in \mathcal{D}^*(\infty)$, we know that $x(t, t_1, y_1)$ exists
and is contained in $\mathcal{D}^*(\infty)$ for any time $t > t_1$. 

Since each function in $\mathcal{F}$ is 
uniformly Lipschitz continuous in $x$ with respect to $t$ on the compact set $\mathcal{D}^*(\infty)$ and
 the number of functions in $\mathcal{F}$ is finite, we can use the Continuous Dependency Theorem of 
initial conditions (\emph{e.g.},~Theorem 3.4 in \cite{khalil2002nonlinear}). 
For $\epsilon > 0$ and $\tilde{t} \geq 0$ there is $\delta(\epsilon, \tilde{t}) >0$ such that 
\begin{align*}
& \|y_1 - y_1'\| \leq \delta \Longrightarrow \|f_{V,m}(x(t_2, t_1, y_1)) - f_{V,m}(x(t_2, t_1, y_1'))\| \leq \epsilon,
\end{align*}
where $t_2 = t_1 + \tilde{t}$.
Let us now choose $\tilde{t} \geq \tilde{t}'(y_1)$ and $\epsilon = \beta(y_1,\tilde{t})/2,$ 
from which it follows that $\epsilon$
is guaranteed to be positive. 
Since $y_1 \in L^+(x_0, t_0)$, there is $t' > t_0$
such that $\|y_1 - x(t',t_0,x_0)\| \leq \delta$. We choose $t_1 = t'$ and
$y_1' = x(t',t_0,x_0)$. But then since $f_{V,m}(x(t_2, t_1, y_1)) \leq \alpha - \beta(y_1, \tilde{t})$ it follows that $f_{V,m}(x(t_2, t_0, x_0)) \leq \alpha -  \beta(y_1, \tilde{t})/2 = \alpha - \epsilon$.
Since $\epsilon > 0$, this contradicts the fact that $\alpha$ is a lower bound for $f_{V,m}$.

Now we shall prove the second part of the statement.
We prove this by a contradiction argument. Suppose there is $\eta > 0$ such that 
there is no $T(\eta) \in \mathbb{R}^+$ such that
$$\min_{t \in [t_0, t_0 + T(\eta)]}\text{dist}(x(t,t_0,x_0), \mathcal{A}) < \eta$$
for all $x_0 \in \mathcal{D}^*(\infty)$ and all $t_0$.
Let $$\beta_{\min} = \min_{z \in \mathcal{D}^*(\infty) \cap \{y : \text{dist}(y, \mathcal{A}) \geq \eta \}}\beta(z,\tilde{t}') > 0.$$
Now, for each positive integer $N$ there is $t_0(N) \geq 0$ and $x_0(N) \in \mathcal{D}^*(\infty)$ such that 
$$\min_{t \in [t_0(N), t_0(N) + N\tilde{t}']}\text{dist}(x(t,t_0(N), x_0(N)), \mathcal{A}) \geq \eta,$$
otherwise we can choose $T(\eta) = N\tilde{t}$, but we assumed that there is no such $T(\eta)$. We have that
\begin{align*}
& f_{V,m}(x(t,t_0(N) + N\tilde{t}',x_0(N))) - f_{V,m}(x(t,t_0(N),x_0(N))) \leq -N\beta_{\min}.
\end{align*}
Now,
\begin{align*}
& f_{V,m}(x(t,t_0(N) + N\tilde{t}',x_0(N))) - f_{V,m}(x(t,t_0(N),x_0(N))) \rightarrow -\infty \quad \text{ as } N \rightarrow \infty,
\end{align*}
which is a contradiction since $f_{V,m}$ is bounded on $\mathcal{D}^*(\infty)$.
\hfill $\blacksquare$

\begin{remark}
Note that the special structure of $\mathcal{A}$ being the consensus set is 
not used in this proof. Also the special structure of $f_{V,m}$ is not used in the proof.
\end{remark}

\begin{lemma}\label{lem:11}
Suppose Assumption \ref{ass:main:0} and \ref{ass:main:1} {(2,3)} hold, $\sigma \in \mathcal{S}_{|\mathcal{F}|,D,U}$, $x^{\sigma}(t_0) \in \mathcal{D}^*(\infty) \cap \mathcal{A}^{c}$ and
$\mathcal{G}_{\sigma(t)}$ is uniformly strongly connected. If $t_0$ is a switching time of $\sigma$, it follows that
$f_{V,m}(x^{\sigma}(t)) < f_{V,m}(x^{\sigma}(t_0))$ for any any $t \geq n(T^{\sigma} + 2\tau_D)$,
where $T^{\sigma}$ is given in Definition~\ref{def:quasi}.
\end{lemma}

\quad \emph{Proof}:   
We assume without loss of generality, 
that the longest time between two consecutive switches of $\sigma(t)$ is bounded from above by 
$2\tau_D$. This assumption is justified by Lemma~\ref{lem:a2}.
Let us consider the solution at an arbitrary switching time $\tau_k$,
and prove that $f_{V,m}(x(n(T^{\sigma} + 2\tau_D) + \tau_k)) < f_{V,m}(x(\tau_k))$. 

\vspace{2mm}
\textbf{Part 1}:\hspace{1mm}
We show that if $i \notin \mathcal{I}_V(\tau_k, s)$, then $i \notin \mathcal{I}_V(\tau_k, t)$ for $t > s \geq \tau_k$.
Suppose that $i \not\in \mathcal{I}_V(\tau_k,s)$ and that there is a $t' >s$ such that $i \in \mathcal{I}_V(\tau_k, t')$. Then since
$V(x_i(t))$ is continuous, there is a $t_1 >s$ such that 
$i \in \mathcal{I}_V(\tau_k, t_1)$ and $i \notin \mathcal{I}_V(\tau_k, t)$ for
$t \in [s, t_1)$. Since $\sigma \in \mathcal{S}_{|\mathcal{F}|, D}$ we know that there is 
$\epsilon > 0$ such that $\sigma(t)$ is constant and $f_i(t,x(t))$ is continuous during $[t_1 -\epsilon, t_1)$,
where $t_1 -\epsilon > s$. 

We define the following constant $$\dot{V}_i^* = \lim_{t \uparrow t_1} \dot{V}(x_i(t)).$$
Now we claim that $\dot{V}_i^* \leq 0$, which we justify as follows. If $t_1$ is not equal to a switching time,
it is immediate that this claim is true since $i \in \mathcal{I}_V(\tau_k, t_1)$, see Assumption \ref{ass:main:1} (2) and Lemma~\ref{lem:dini}.
On the other hand, if $t_1$ is equal to a switching time, the claim is also true
and can be shown as follows.
If $\sigma$ is the switching signal function for our solution, we can create 
another switching signal function $\sigma' \in \mathcal{S}_{|\mathcal{F}|,D}$ which satisfies
$$\sigma'(t) = \sigma(t) \: \: \: 0 \leq t < t_1 \quad \text{ and } \quad  \sigma'(t_1) = \sigma(t_1-\epsilon).$$
So,
\begin{align*}
\dot{V}_i^* & = \lim_{t \uparrow t_1} \dot{V}(x_i(t)) =  \dot{V}(x_i^{\sigma'}(t)) \leq 0,
\end{align*}
where the last inequality follows from Assumption \ref{ass:main:1} (2) and Lemma~\ref{lem:dini}.

We now know that $\dot{V}_i^* \leq 0$.
Thus there are two options for $\dot{V}^*$; either it is (1) strictly negative or (2) zero.
In case (1), since $\sigma(t)$ is piecewise right-continuous there is a positive  $\epsilon' < \epsilon$ such that
$\dot{V}(x_i(t))$ is continuous and strictly negative on $[t_1 -\epsilon', t_1)$. We also know, since $V(x_i(t_1)) = f_{V,m}(x(\tau_k))$, that $V(x_i(t)) \leq V(x_i(t_1))$
for all $t \geq \tau_k$. Using these two facts, we get that
$$V(x_i(t_1)) = V(x_i(t_1 - \epsilon')) + \int_{t_1 - \epsilon'}^{t_1}\dot{V}(x_i(t))dt < V(x_i(t_1))$$
which is a contradiction. 

Now we consider case (2).
By using Assumption \ref{ass:main:1} (3) we can show that
$$x(t_1) = \lim_{t \uparrow t_1} x(t)$$
satisfies $x_i(t_1) = x_j(t_1)$  and
$\lim_{t \uparrow t_1} \dot{V}(x_j(t)) = 0$ for all $j \in \mathcal{N}_i(t_1 -\epsilon)$
(note that $\sigma(t)$ is constant on $[t_1 - \epsilon, t_1)$, so $\mathcal{N}_i(t) = \mathcal{N}_i(t_1 -\epsilon)$
on this half-open interval), otherwise $V(x_j(t_1)) = f_{V,m}(x(\tau_k))$ and 
$$\lim_{t \uparrow t_1} \dot{V}(x_j(t)) < 0,$$
which we just showed is a contradiction.
For any $j$ such that $j \in \mathcal{N}_i(t_1 -\epsilon)$ it holds that $x_k(t_1) = x_j(t_1)$  and
$\lim_{t \uparrow t_1} \dot{V}(x_k(t)) = 0,$
for all $k \in \mathcal{N}_j(t_1 -\epsilon)$.
By using the same argument for the 
neighbors of the neighbors of agents in $\mathcal{N}_i(t_1 - \epsilon)$ and so on, we get that
$x_i(t_1) = x_j(t_1)$ for all $j$ that belongs to the connected component of node $i$ in 
$\mathcal{G}_{\sigma(t_1 - \epsilon)}$. Let us denote the state in this connected
component by $x_{c_i}(t)$, where $c_i \subset \mathcal{V}$ are all neighbors in this connected component. 
It holds that $$\lim_{t \uparrow t_1} \dot{V}(x_j(t)) = 0,$$
for all $j \in c_i$.
During $[t_1- \epsilon, t_1)$ the dynamics for $x_{c_i}$ is
$$\dot{x}_{c_i} = f^{c_i}(t,x_{c_i}).$$
The function $f^{c_i}$ is the part of ${f}$
corresponding to the connected component $c_i$. 
By using Assumption~\ref{ass:main:1} (3) we get that 
$$\lim_{t \uparrow t_1} f^{c_i}(t,x_{c_i}(t)) = 0,$$
which is a contradiction, since $x_{c_i}$ cannot reach such an
equilibrium point in finite time without violating the uniqueness of the solution property
(the functions in $\mathcal{F}$ are continuous in $t$ and Lipschitz in $x$).

\vspace{2mm}
\textbf{Part 2}: \hspace{1mm} 
Using part 1 we show that $\mathcal{I}_V(\tau_k, t)$ is empty for $t \geq n(T^{\sigma} + 2\tau_D) + \tau_k$.
Suppose that $\mathcal{I}(\tau_k, \tau_{k}) \subset \mathcal{I}(\tau_k, \tau_{k'})$, where $\tau_{k'}$ is the 
first switching time after $\tau_k + T^{\sigma}$.  We know from part 1 that $\mathcal{I}(\tau_k, \tau_{k})^c \subset \mathcal{I}(\tau_k, \tau_{k'})^c$ 
(where complements are taken with respect to the set $\mathcal{V}$)
which implies that $\mathcal{I}(\tau_k, \tau_{k'}) \subset \mathcal{I}(\tau_k, \tau_{k})$, so our assumption has the consequence that
$\mathcal{I}(\tau_k, \tau_{k}) = \mathcal{I}(\tau_k, \tau_{k'})$. Now, since $\mathcal{G}_{\sigma(t)}$ is uniformly 
strongly connected, there is a switching time $\tau_{k''}$ such that $\tau_{k} \leq \tau_{k''} \leq \tau_{k} + T^{\sigma}$ for 
which there are $i,j$ that satisfy $i \in \mathcal{I}(\tau_k, \tau_{k})$, $j \in \mathcal{I}(\tau_k, \tau_{k})^c$ and $j \in \mathcal{N}_i(\tau_{k''})$. But then 
$j \in \mathcal{N}_i(s)$ for $s \in [\tau_{k''}, \tau_{k''} + \tau_D)$. Thus, $i \in \mathcal{I}^*_V(s)$ for $s \in [\tau_{k''}, \tau_{k''} + \tau_D)$,
which means that $\dot{V}(x_i(s)) <0$ on $[\tau_{k''}, \tau_{k''} + \tau_D)$. But since $i \in \mathcal{I}_V(\tau_k,s)$ for $s \in [\tau_{k''}, \tau_{k''} + \tau_D)$, the function ${V}(x_i(s))$ is constant on $[\tau_{k''}, \tau_{k''} + \tau_D)$, which is a contradiction.
Our hypothesis that $\mathcal{I}(\tau_k, \tau_{k}) \subset \mathcal{I}(\tau_k, \tau_{k'})$ leads to a contradiction. 
Thus, $\mathcal{I}(\tau_k, \tau_{k'})$ is a strict subset 
of $\mathcal{I}(\tau_k, \tau_{k})$.

Now, there are two cases for $\mathcal{I}_V(\tau_{k}, \tau_{k'})$. It is either (1) empty,
or {(2)} nonempty. In case {(1)} we are done. In case (2) 
we have that $\mathcal{I}_V(\tau_k, \tau_{k'}) = \mathcal{I}_V(\tau_{k'}, \tau_{k'})$.
We know that $\tau_{k'} \leq \tau_k + T^{\sigma} + 2\tau_D$ by the assumption that $\tau_U = 2\tau_D$. Now we can apply the same procedure
for the set $\mathcal{I}_V(\tau_{k'}, \tau_{k'})$. By repeating the procedure $n$ times, we know that $\mathcal{I}_V(\tau_k, t) = \emptyset$ for $t  \geq n(T^{\sigma} + 2\tau_D) + \tau_k$. \hfill $\blacksquare$

\vspace{3mm}
\quad \emph{Proof of Theorem \ref{chapter1:thm:3}}:   
We prove this theorem by showing that there is a function $\beta$ with
the properties given in Lemma \ref{lem:11}. For each $\sigma \in \mathcal{S}_{|\mathcal{F}|,D,U}$,
there is a corresponding $\beta$.

Initially we assume that $t_0$ is a switching time. 
This assumption will be relaxed towards the end of the proof, so that we consider arbitrary times. 
We assume once again without loss of generality that $\tau_U = 2\tau_D$,
and from Lemma \ref{lem:11} it follows that for a switching time $t_0$, it holds that $f_{V,m}(x(t_0 + \tilde{t})) < f_{V,m}(x(t_0))$
where $\tilde{t} \geq n(T^{\sigma}+2\tau_D)$. In the following, let us choose $\tilde{t} \geq \tilde{t}' =  n(2T^{\sigma}+2\tau_D)$. Obviously, 
since $f_{V,m}(x(t))$ is decreasing,
$f_{V,m}(x(t_0 + \tilde{t})) < f_{V,m}(x(t_0))$ for $\tilde{t} \geq \tilde{t}'$, and this particular choice of
$\tilde{t}'$ will have its explanation towards the end of the proof. 

During the time interval $[t_0, t_0 + \tilde{t}]$ there is  an upper bound $M_u$ 
and a lower bound $M_d$ 
on the number of
switches of $\sigma(t)$. 
 Now we create something which we call scenarios.
A scenario $s$ is defined as follows,
 $$s = ({f}_0', {f}'_1, \ldots {f}_k').$$
The function ${f}_i' \in \mathcal{F}$ for $i \in \{1, \ldots, k \}$, where $k \in \{M_d, M_d +1, \ldots, M_u\}$. What this illustrates is that during the time period between $t_0$ and the first switching time $\tau_1$ after $t_0$,
the function $f'_0$ is the right-hand side of \eqref{chapter1:dynamics}, during the second time period between $\tau_1$ and 
$\tau_2$, $f'_1$ is the right-hand side of \eqref{chapter1:dynamics}
and so on. 
By a slight abuse of notation, $\tau_1$ is the first switching time after $t_0$ and  $\tau_i$ is the first switching time after $\tau_{i-1}$
for $i \in \{2, \ldots, k \}$.
The number of possible scenarios is finite and do not dependent on where the actual switches occur in time. 

Now, for a specific scenario $s$ with $k$ switching times, and where the switching times are the elements in the vector
$\tau = (\tau_1, \ldots, \tau_k)^T$, we write the solution to \eqref{chapter1:dynamics} as
\begin{align*}
& x^{(s,\tau)}(t_0 + \tilde{t}) = 
 x^{(s,\tau)}(t_0) + \int_{t_0}^{\tau_1}f'_0(t-t_0,x^{(s,\tau)}(t))dt +  \ldots \\
& +\int_{\tau_{k-1}}^{\tau_k}f'_{k-1}(t-\tau_{k-1},x^{(s,\tau)}(t))dt  
+ \int_{\tau_{k}}^{t_0 + \tilde{t}}f'_k(t-\tau_{k},x^{(s,\tau)}(t)).
\end{align*}
Thus, instead of parameterizing $x$ by the switching signals, we here on the interval $[t_0, t_0 + \tilde{t}]$ parameterize $x$ by the
scenarios and the switching times vector $\tau$. 

The function $x^{(s,\tau)}(t_0 + \tilde{t})$
is continuous in $\tau$ on the set 
 \begin{align*}
\mathcal{C}_s = \{ \tau :\;
& {t}_0 \leq \tau_i \leq t_0 + \tilde{t} \text{ for } i = 1,\ldots, k,\\
&\tau_1 \geq t_0 + \tau_D, \\
&\tau_1 \leq t_0 + 2\tau_D, \\
&{ \tau}_{i+1} \geq \tau_{i} + \tau_D \text{ for } i = 1,\ldots, k-1, \\
&{ \tau}_{i+1} \leq \tau_{i} + 2\tau_D \text{ for } i = 1,\ldots, k-1 \\
& t_0 + \tilde{t} \leq \tau_k + 2\tau_D \}.
\end{align*}
This is a consequence of the Continuous Dependency Theorem of initial conditions and is shown by 
the following argument. For a specific $\tau$, suppose
 $\tau_i$ is changed
to $\tau_i'$, where $|\tau_i' - \tau_i|$ is small and $i \in \{1, \ldots, k\}$. Then we define 
$\tau' = (\tau_1, \ldots, \tau_{i-1}, \tau_i', \tau_{i +1}, \ldots, \tau_k)^T$.
\begin{align*}
& {x}^{(s,\tau')}(t_0 + \tilde{t}) = x^{(s,\tau')}(t_0) + \int_{t_0}^{\tau_1}f'_0(t-t_0,x^{(s,\tau')}(t))dt + \\ 
& \ldots +\int_{\tau_{i-1}}^{\tau_i'}f'_{i-1}(t-\tau_{i-1},x^{(s,\tau')}(t))dt + \int_{\tau_{i}'}^{\tau_{i+1}}f'_i(t -\tau_{i}' ,x^{(s,\tau')}(t))dt  + \\
& \ldots + \int_{\tau_{k}}^{t_0 + \tilde{t}}f'_k(t-\tau_{k},x^{(s,\tau')}(t)),
\end{align*}
so ${x}^{(s,\tau')}$ is an alternative solution where $\tau_i$ is replaced by $\tau_i'$. 
We know that all such alternative solutions exist and $x^{(s,\tau')}(t) \in \mathcal{D}^{*}(\infty)$ for $t \in [t_0, t_0+ \tilde{t}]$.

Now,
\begin{align*}
x^{(s,\tau)}(t_0 + \tilde{t}) & = x^{(s,\tau)}(\tau_i) + \int_{\tau_{i}}^{\tau_{i+1}}f'_i(t-\tau_i,x^{(s,\tau)}(t))dt + \\
& \ldots + \int_{\tau_{k}}^{t_0 + \tilde{t}}f'_k(t-\tau_{k},x^{(s,\tau)}(t)), \\
{x}^{(s,\tau')}(t_0 + \tilde{t}) & = x^{(s,\tau')}(\tau_i') + \int_{\tau_{i}'}^{\tau_{i+1}}f'_i(t-\tau_i',x^{(s,\tau')}(t))dt + \\
& \ldots + \int_{\tau_{k}}^{t_0 + \tilde{t}}f'_k(t-\tau_{k},x^{(s,\tau')}(t)).
\end{align*}
As $|\tau_i - \tau_i'| \rightarrow 0$ it holds that 
$$\|x^{(s,\tau)}(\tau_{i+1},\tau_i,x^{(s,\tau)}(\tau_i)) - x^{(s,\tau')}(\tau_{i+1},\tau_i',x^{(s,\tau')}(\tau_i'))\| \rightarrow 0,$$
which implies that
\begin{align*}
& \|x^{(s,\tau)}(t_0 + \tilde{t},\tau_{i+1},x^{(s,\tau)}(\tau_{i+1},\tau_i,x^{(s,\tau)}(\tau_i))) \\
&- x^{(s,\tau')}(t_0 + \tilde{t},\tau_{i+1},x^{(s,\tau')}(\tau_{i+1},\tau_i',x^{(s,\tau')}(\tau_i')))\| \rightarrow 0.
\end{align*}
The function $f_{V,m}(x^{(s,\tau)}(t_0 + \tilde{t}, t_0, x_0))$ is 
also continuous in $\tau$ on $\mathcal{C}_s$.

Only a subset of all scenarios are feasible.
We say that a scenario is feasible if there is $\tau' \in \mathcal{C}_s$
and a switching signal function $\sigma'$ such that $T^{\sigma'} = T^{\sigma}$ and
where $x^{\sigma'}(t) = x^{(s,\tau')}(t)$ for $t \in [t_0, t_0 + \tilde{t}']$.
According to Lemma~\ref{lem:11}, this means that  
$f_{V,m}(x(t_0)) -f_{V,m}(x^{(s,\tau)}(t_0 + \tilde{t}', t_0, x_0))> 0$
for the $\tau' \in \mathcal{C}_s$.
Now, suppose the scenario $s$ is feasible, the question
is if it is true that
$$f_{V,m}(x(t_0)) -f_{V,m}(x^{(s,\tau)}(t_0 + \tilde{t}', t_0, x_0))> 0$$
for all $\tau \in \mathcal{C}_s$. 
By the subsequent argument we show that this is true.

Suppose $s$ is feasible, then there is $\tau \in \mathcal{C}_s$ such that there is
a switching signal function $\sigma'$ (not necessarily $\sigma$)
which has switching times equal to the elements in $\tau$ during $[t_0, t_0 + \tilde{t}']$
and $x^{\sigma'}(t) = x^{(s,\tau)}(t)$ for $t \in [t_0, t_0 + \tilde{t}']$.
The graph $\mathcal{G}_{\sigma'(t)}$ is uniformly strongly connected and 
 $T^{\sigma} = T^{\sigma'}$. Now, if the elements in $\tau$ are changed by means of a
continuous transformation 
to an arbitrary $\tau'' \in \mathcal{C}_s$, then there is a $\sigma'' \in \mathcal{S}_{|\mathcal{F}|,D,U}$
for which $\mathcal{G}_{\sigma''(t)}$ is uniformly strongly connected.
The switching times of $\sigma''$ are given by the elements in $\tau''$ during $[t_0,t_0+ \tilde{t}']$, and an upper bound on the length of an half-open interval in time such that 
the union graph $\mathcal{G}_{\sigma''(t)}$ is strongly connected during that interval is $T^{\sigma''}= 2T^{\sigma}$.
This is true since we know that the lower bound between two switching times is $\tau_D$ and the 
upper bound is $2\tau_D$. Thus, by changing $\tau$ to $\tau''$, the length of any interval between two consecutive switching times can at most be changed to be twice
as long.
Now, according to Lemma~\ref{lem:11}, since $\mathcal{G}_{\sigma''(t)}$ is uniformly 
 strongly connected (with an upper bound of $2T^{\sigma}$ on the length of the interval such that 
the union graph is strongly connected) we know that since $\tilde{t}' = n(2T^{\sigma'} + 2\tau_D) = n(T^{\sigma''} + 2\tau_D)$,
$$f_{V,m}(x(t_0)) -f_{V,m}(x^{(s,\tau'')}(t_0 + \tilde{t}', t_0, x_0))> 0.$$ 
Because $\tau''$ is arbitrary in $\mathcal{C}_s$, if $s$ is feasible it holds that
$$f_{V,m}(x(t_0)) -f_{V,m}(x^{(s,\tau)}(t_0 + \tilde{t}', t_0, x_0))> 0$$
for all $\tau \in \mathcal{C}_s$.

By choosing $\tilde{t} \geq \tilde{t}' $, we now know that for feasible $s$ it holds that
$$f_{V,m}(x^{(s,\tau)}(t_0 + \tilde{t}, t_0, x_0)) - f_{V,m}(x_0) < 0$$
for all $\tau$ in $\mathcal{C}_s$.
By Weierstrass Extreme Value Theorem there exists $\tau^* \in \mathcal{C}_s$
such that
\begin{align*}
\delta_{s}(x_0,\tilde{t}) & = \min_{\tau \in \mathcal{C}_s}f_{V,m}(x_0)-f_{V,m}(x^{(s,\tau)}(t_0 + \tilde{t}, t_0, x_0)) \\
& = f_{V,m}(x(t_0)) -f_{V,m}(x^{(s,\tau^*)}(t_0 + \tilde{t}, t_0, x_0))> 0.
\end{align*}
Note that this $\delta_s$ is not a function of $t_0$,
since all possible switching signal functions  are accounted for during $[t_0, t_0 + \tilde{t}]$ for the specific
scenario. Thus, $t_0$ could be any switching time of $\sigma$.

Now,
\begin{align*}
 \inf_{t_0 \in \{\tau_k\}} f_{V,m}(x_0) -f_{V,m}(x^{\sigma}(t_0 + \tilde{t}, t_0, x_0)) \geq & \\
\min_{s}\min_{\tau \in \mathcal{C}_s}f_{V,m}(x_0)-f_{V,m}(x^{(s,\tau)}(t_0 + \tilde{t}, t_0, x_0)) =&  \\
 \min_{s}\delta_s(x_0,\tilde{t}) >& 0,
\end{align*}
where $\{\tau_k\}$ is the set of all switching times of $\sigma$.
The set of scenarios that we minimize over are only feasible scenarios.
 Now we define $$\beta(x_0,\tilde{t}) = \min_{s}\delta_s(x_0,\tilde{t} - 2\tau_D),$$
where $\delta_s$ is defined as zero for negative second arguments. The subtraction by $2\tau_D$ is due to the fact that 
$t_0$ was assumed to be a switching time, hence we subtract this term in order to be sure that $-\beta(x_0,\tilde{t})$
does not overestimate the decrease of $f_{V,m}(x(t))$.

Now we need to prove that $\beta(x_0,\tilde{t})$ is lower semi-continuous in $x_0$.
We show that
$\delta_s(x_0,\tilde{t})$ is continuous in $x_0$ for all $s$. From this fact it follows that $\beta(x_0, \tilde{t})$
is continuous in $x_0$. The function 
$$g_s(\tau, \tilde{t}, x_0) =  f_{V,m}(x_0) -f_{V,m}(x^{(s,\tau)}(t_0 + \tilde{t},t_0,x_0))$$
is continuous in $\tau$ and $x_0$. 
It follows directly that $\delta_s$ is continuous in $x_0$, since 
\begin{equation}\label{olle111}
\delta_{s}(x_0,\tilde{t})  = \min_{\tau \in \mathcal{C}_s}g_s(\tau, \tilde{t}, x_0)
\end{equation}
and $\mathcal{C}_s$ is compact.
\hfill $\blacksquare$

Now we turn to the proof of Theorem \ref{chapter1:thm:5}, but first
we formulate some lemmas necessary in order to prove this theorem. 
Before we proceed, let us define $$\bar{B}_{r,mn}(\mathcal{A}) = \{x \in \mathbb{R}^{mn}: \textnormal{dist}(x,\mathcal{A}) \leq r\}.$$

\begin{lemma}\label{lem:10}
Suppose $V$ fulfills Assumption \ref{ass:main:2} (1),
then  for $x \in \bar{B}_{r,mn}(\mathcal{A}) \cap \mathcal{D}$
there are class $\mathcal{K}$ functions $\beta_1$ and $\beta_2$ on $[0,r]$ such that 
$$\beta_1(\textnormal{dist}(x,\mathcal{A})) \leq f_{V,m,m}(x) \leq \beta_2(\textnormal{dist}(x,\mathcal{A})).$$
\end{lemma}

\vspace{3mm}
\quad \emph{Proof}: 
We follow the procedure in the proof of Lemma 4.3 in \cite{khalil2002nonlinear}
and define 
$$\psi(s) = \inf_{\{s \leq \textnormal{dist}(x,\mathcal{A}) \leq r\} \cap \mathcal{D}}f_{V,m,m}(x) \quad \text{for }  0 \leq s \leq r$$
from which we have that $\psi(\textnormal{dist}(x,\mathcal{A})) \leq f_{V,m,m}(x)$ on $B_{r,mn}(\mathcal{A}) \cap \mathcal{D}$.
We also define 
$$\phi(s) = \sup_{\{\textnormal{dist}(x,\mathcal{A}) \leq s\} \cap \mathcal{D}} f_{V,m,m}(x) \quad \text{for } 0 \leq s \leq r$$
from which we have that $f_{V,m,m}(x) \leq \phi(\textnormal{dist}(x,\mathcal{A}))$ on $B_{r,mn}(\mathcal{A}) \cap \mathcal{D}$.
The functions $\psi$ and $\phi$ are continuous, positive definite and 
increasing, however not necessarily strictly increasing. The positive definiteness of $\psi$
is guaranteed by the fact that $\inf$ is taken over compact
sets, and since $f_{V,m,m}(x)$ is positive and continuous on the sets
the result follows by using Weierstrass Extreme Value Theorem.

Now there exist class $\mathcal{K}$ functions $\beta_1$ and $\beta_2$ such that 
$\beta_1(s) \leq k\psi(s)$ for some $k \in (0,1)$, and $\beta_2(s) \geq k\phi(s)$ for
some $k > 1$ where $s \in [0,r]$. It follows that
$$\beta_1(\textnormal{dist}(x,\mathcal{A})) \leq f_{V,m,m}(x) \leq \beta_2(\textnormal{dist}(x,\mathcal{A}))$$
on $\bar{B}_{r,mn}(\mathcal{A}) \cap \mathcal{D}$. 
\hfill $\blacksquare$

\begin{lemma}\label{thm:2}
Suppose $x(t) \in \mathcal{D}$ for all $t \geq t_0$ and Assumption \ref{ass:main:2} (1,2) holds,
then the set $\mathcal{A}$ is uniformly stable for \eqref{chapter1:dynamics}.
\end{lemma}

\vspace{3mm}
\quad \emph{Proof}:   
Compared to the proof of Theorem \ref{chapter1:thm:1} we do not have to address
the issue of existence of the solution, since by assumption it
exists in $\mathcal{D}$. Using Assumption \ref{ass:main:2} (2) we get from the 
Comparison Lemma (\emph{e.g.}, Lemma 3.4 in~\cite{khalil2002nonlinear}),
that $$f_{V,m,m}({x}(t)) \leq f_{V,m,m}(x_0).$$

From Lemma \ref{lem:10} we know that there exist class $\mathcal{K}$ functions $\beta_1$ and $\beta_2$ defined on $[0,r]$ such that
$$\beta_1(\textnormal{dist}(x,\mathcal{A})) \leq f_{V,m,m}(x) \leq \beta_2(\textnormal{dist}(x,\mathcal{A})).$$
Now let $\epsilon \in (0,r)$ and $\delta = \beta_2^{-1}(\beta_1(\epsilon))$. Then 
if $x(t_0) \in B_{\delta,mn}(\mathcal{A})$, it follows that
\begin{align*}
\textnormal{dist}(x,\mathcal{A}) & \leq \beta_1^{-1}(f_{V,m,m}(x(t))) \leq  \beta_1^{-1}(f_{V,m,m}(x_0)) \\
& \leq \beta_1^{-1}(\beta_2(\textnormal{dist}(x(t_0),\mathcal{A}))  \leq \beta_1^{-1}(\beta_2(\delta)) = \epsilon.
\end{align*} \hfill $\blacksquare$
\\ \\
If $x_0 \in \mathcal{D}^{*}(\infty)$, the set $\mathcal{A}$ is uniformly stable
for any $\sigma \in \mathcal{S}_{|\mathcal{F}|,D}$.

\begin{lemma}\label{lem:13}
Suppose $x_0 \in \mathcal{A}^c  \cap \mathcal{D}^*(\infty)$ and $t_0$ are arbitrary and
Assumption \ref{ass:main:2} (1,2) holds.
Suppose
there is a non-negative function 
$$\beta(y, \tilde{t}): \mathbb{R}^+ \times \mathbb{R}^+ \rightarrow \mathbb{R}^+$$
that is increasing in $\tilde{t}$
and lower semi-continuous in $y$.
Furthermore, suppose there is $\tilde{t}' > 0$, such that
for $\tilde{t} \geq \tilde{t}'$, it holds that $\beta(y, \tilde{t}) >0$
for all $y \in \mathbb{R}^{++}$.

If
\begin{align*}
f_{V,m,m}(x(t,t_0, x_0)) - f_{V,m,m}(x_0)
\leq -\beta(\textnormal{dist}(x_0, \mathcal{A}), t-t_0),
\end{align*}
$\mathcal{A}$ is globally uniformly asymptotically stable relative to $\mathcal{D}^*(\infty)$.
\end{lemma}

\vspace{3mm}
\quad \emph{Proof}:  
We already know from Lemma \ref{thm:2} that $\mathcal{A}$ 
is uniformly stable relative to $\mathcal{D}^*(\infty)$. What is left to
prove is that
$\mathcal{A}$ is globally uniformly attractive relative to $\mathcal{D}^*(\infty)$. 
In order to show this, the procedure is 
analogous to the procedure in Lemma \ref{lem:11}, where we use the positive
limit set $L^+(x_0, t_0)$ for the solution $x(t, t_0, x_0)$. 

Let us consider arbitrary $t_0$ and $x_0 \in \mathcal{D}^*(\infty) \cap \mathcal{A}^c$. 
By using the fact that $f_{V,m,m}(x(t))$ is continuous and $\mathcal{D}^*(\infty)$ is compact and invariant, it follows that $f_{V,m,m}(x(t))$
converges to a lower bound $\alpha(x_0,t_0) \geq 0$ as $t\rightarrow \infty$. 
Suppose that $L^+(x_0, t_0) \not \subset \mathcal{A}$. We want to prove that $\mathcal{A}$ is attractive by 
showing that this assumption leads to a contradiction. Let $t_1 = t_0 + \tilde{t}'$
and let $y_1$ be an arbitrary point in $L^+(x_0, t_0) \cap \mathcal{A}^c \subset \mathcal{D}^{*}(\infty)$.
 By using the Continuous Dependency Theorem of 
initial conditions (\emph{e.g.}~Theorem 3.4 in \cite{khalil2002nonlinear}),
for any $\epsilon > 0$ there is $\delta(\epsilon, \tilde{t}') >0$ such that 
\begin{align*}
\|y_1 - y_1'\| \leq \delta \Longrightarrow  
\|f_{V,m,m}(x(t_2, t_1, y_1)) - f_{V,m,m}(x(t_2, t_1, y_1'))\| \leq \epsilon,
\end{align*}
where $t_2 = t_1 + \tilde{t}'$.
Let us choose $\epsilon = \beta(\textnormal{dist}(y_1, \mathcal{A}), \tilde{t}')/2$. 
Since $y_1 \in L^+(x_0, t_0)$, there is a $t'$
such that $\|y_1 - x(t',t_0,x_0)\| \leq \delta$, thus we choose $t_1 = t'$ and
$y_1' = x(t',t_0,x_0)$. But then $$f_{V,m,m}(x(t_2, t_0, x_0)) \leq \alpha -  \beta(\textnormal{dist}(y_1, \mathcal{A}), \tilde{t}')/2 < \alpha,$$
which contradicts the fact that $\alpha$ is a lower bound for $V$.
Hence, $x(t,t_0,x_0) \rightarrow \mathcal{A}$ as $t \rightarrow \infty$ for all $t_0$ and $x_0 \in \mathcal{D}^*(\infty)$.

What is left to prove is that for all $ \eta >0$ and $x_0 \in \mathcal{D}^*(\infty)$, there is $T(\eta)$ such
that $$t \geq t_0 + T(\eta) \Longrightarrow \text{dist}(x(t, t_0,x_0), \mathcal{A}) < \eta.$$

We use a contradiction argument. Suppose there is an $\eta >0$ such that there is no such $T(\eta)$. 
We know, since $\mathcal{A}$ is uniformly stable relative to $\mathcal{D}^*(\infty)$, that there is a $\delta'(\eta) > 0$ such that 
for $x_0 \in \mathcal{D}^*(\infty)$ it holds that $$\textnormal{dist}(x_0,\mathcal{A}) \leq \delta' \Longrightarrow \textnormal{dist}(x(t),\mathcal{A}) \leq \eta$$
for all $t \geq t_0$.
Let $$d_{\max} = \max\limits_{y \in \mathcal{D}^{*}(\infty)}\textnormal{dist}(y,\mathcal{A})$$
and  $$\beta' = \min\limits_{d \in [\delta'(\eta), d_{\max}]}\beta(d, \tilde{t}') > 0.$$
For any (positive integer) $N$ there are $t_0(N)$ and $x_0(N)$ in $\mathcal{D}^*(\infty)$ such that 
$$\textnormal{dist}(x(t,t_0(N),x_0(N)),\mathcal{A}) > \delta'$$ 
 when $t_0 \leq t \leq t_0 + N\tilde{t}'$, otherwise $T(\eta)$ would exist which we assume it does not.
From this it follows that 
\begin{align*}
f_{V,m,m}(x(t_0(N) + N\tilde{t}',t_0(N),x_0(N))) 
- f_{V,m,m}(x_0(N)) \leq -N\beta'. 
\end{align*}
Since $\beta'$ is a constant, it follows that 
\begin{align*}
\lim_{N \rightarrow \infty}(f_{V,m,m}(x(t_0(N) + N\tilde{t}',t_0(N),x_0(N))) 
-f_{V,m,m}(x_0(N))) - \infty. 
\end{align*}
This is a contradiction 
since $f_{V,m,m}$ is bounded on $\mathcal{D}^{*}(\infty)$.
\hfill $\blacksquare$

\begin{lemma}\label{lem:21}
Suppose Assumption \ref{ass:main:0} and \ref{ass:main:2} (1,2,3) hold, $x_0 \in \mathcal{D}^*(\infty) \cap \mathcal{A}^c$ and $\sigma \in \mathcal{S}_{|\mathcal{F}|,D,U}$. Furthermore, suppose 
$\mathcal{G}_{\sigma(t)}$ is uniformly quasi-strongly connected,
then $$f_{V,m,m}(x^{\sigma}(t)) < f_{V,m,m}(x_0)$$ if $t_0$ is a switching time and $t \geq n(T^{\sigma} + 2\tau_D) + t_0$,
where $T^{\sigma}$ is given in Definition~\ref{def:quasi}.
\end{lemma}

\vspace{3mm}
\quad \emph{Proof}: 
The proof of this lemma is to a large extent similar to the proof
of Lemma \ref{lem:11} and hence omitted.
In part 1, instead of one connected component $c_i$, there are two connected components,
where the states in the connected components reach an equilibrium in finite time which cannot 
be reached since the right-hand side of the dynamics is Lipschitz in $x$. Thus, one obtains the desired 
contradiction.
In part 2, the main difference is that now
$$\mathcal{J}_V(\tau_k, \tau_k + T^{\sigma} + 2\tau_D)$$
is a strict subset of $\mathcal{J}_V(\tau_k, \tau_k)$ and the graph is uniformly quasi-strongly connected
instead of uniformly strongly connected.
The reason for 
not letting the graph be uniformly quasi-strongly connected in Lemma \ref{lem:11}, is that if
it is uniformly quasi-strongly connected, we might have the situation that
the union graph during $[\tau_k, \tau_k + T^{\sigma})$ is a rooted spanning tree, with the root corresponding to an 
agent in $\mathcal{I}_V(\tau_k, \tau_k)$
and in that case $\mathcal{I}_{V}(\tau_k, \tau_k) = \mathcal{I}_{V}(\tau_k, \tau_k + T^{\sigma} + 2\tau_D)$ might hold.
\hfill $\blacksquare$

\vspace{3mm}
\quad \emph{Proof of Theorem \ref{chapter1:thm:5}}:   
\textbf{Only if}: 
Assume $\mathcal{G}_{\sigma(t)}$ is not uniformly quasi-strongly connected. 
Then for any $T' > 0$ there is $t_0(T')$ such that the union graph $\mathcal{G}([t_0, t_0 + T'))$ is not quasi 
strongly connected. During $[t_0, t_0 + T')$ the set of nodes $\mathcal{V}$ can be divided into
two disjoint sets of nodes $\mathcal{V}_1$ and $\mathcal{V}_1$ (see proof of Theorem 3.8 in \cite{lin2007state})
where there are no edges $(i,j)$ or $(j,i)$ in $\mathcal{G}([t_0, t_0 + T'))$ such that $i \in \mathcal{V}_1$ and 
 $j \in \mathcal{V}_2$ or $j \in \mathcal{V}_1$ and 
 $i \in \mathcal{V}_2$ respectively.

We introduce $y_1^*, y_2^* \in \mathcal{D}^*(\infty)$, where $y_1^* \neq y_2^*$ and
let $x_i(t_0) = y_1^*$ and $x_j(t_0) = y_2^*$ for all $i \in \mathcal{V}_1$, $j \in \mathcal{V}_2$. 
Let $\eta = \text{dist}(x_0, \mathcal{A})/2$. Suppose now that $\mathcal{A}$ is globally uniformly asymptotically stable
relative to $\mathcal{D}^*(\infty)$,
then there is a $T(\eta)$ such that
$$t \geq t_0 + T(\eta) \Longrightarrow \text{dist}(x(t), \mathcal{A}) < \eta.$$
We choose $T' > T(\eta)$. Due to 
Assumption \ref{ass:main:2} (3) we have that $x_i(t) = y_1^*$ and $x_j(t) = y_2^*$
when $i \in \mathcal{V}_1$ and $j \in \mathcal{V}_2$ for $t \in [t_0(T'), t_0(T') +T')$. 
Thus, $\text{dist}(x(t), \mathcal{A}) > \eta$ for some $t \geq t_0(T') + T(\eta)$ which is a contradiction. 
\newline

\noindent
\textbf{If}:
Once again we assume without loss of generality that $\tau_U = 2\tau_D$.
We prove this part of the proof by constructing a function $\beta$ according to Lemma \ref{lem:13}.
The proof is to a large extent similar to the proof of Theorem~\ref{chapter1:thm:3} and
hence only the important part is addressed.
Along the lines of the proof of  Theorem~\ref{chapter1:thm:3},
we define $\delta_s(x_0, \tilde{t})$,
where we use Lemma~\ref{lem:21}
which assures that if $t_0$ is a switching time of $\sigma$ and $\tilde{t}' = n(2T^{\sigma} + 2\tau_D)$,
it holds that $$f_{V,m,m}(x(t_0 + \tilde{t}')) < f_{V,m,m}(x(t_0))$$
for $x_0 \in \mathcal{A}^c \cap \mathcal{D}^{*}(\infty).$

Now we define 
\begin{align*}
\beta(v, \tilde{t}) =& \min_s\min_{\mathcal{D}^{*}(\infty) \cap \{x_0 :\text{dist}(x_0, \mathcal{A}) = v\}} (\delta_s(x_0, \tilde{t} - 2\tau_D)),
\end{align*}
where the minimization is over feasible scenarios only. Feasible scenarios are defined
in the analogous way as in the proof of Theorem~\ref{chapter1:thm:3}. 
Since $\mathcal{D}^{*}(\infty) \cap \{x_0 :\text{dist}(x_0, \mathcal{A}) = v\}$ is compact
and $\delta_s(x_0, \tilde{t})$ is positive and continuous on this set for $\tilde{t} \geq \tilde{t}'$, it holds that
$\beta(v, \tilde{t})$ is positive for positive $v$. Also $\beta(v, \tilde{t})$ is actually not only lower semi-continuous, but continuous in $v$.
\hfill $\blacksquare$
\vspace{2mm}

Note, that in the \textbf{only if} part of the proof of Theorem \ref{chapter1:thm:5} we have not shown that $x(t) \not\rightarrow \mathcal{A}$
when $t \rightarrow \infty$ if $\mathcal{G}_{\sigma(t)}$ is not uniformly quasi-strongly connected. But
we can guarantee that if convergence would occur, it cannot be uniform if $\mathcal{G}_{\sigma(t)}$ is not uniformly quasi-strongly connected

\bibliographystyle{unsrt}        
\bibliography{arxiv_thunberg1}                                 

\begin{thebibliography}{10}

\bibitem{aeyels1995asymptotic}
D.~Aeyels.
\newblock Asymptotic stability of nonautonomous systems by liapunov's direct
  method.
\newblock {\em Systems \& Control Letters}, 25(4):273--280, 1995.

\bibitem{afsari2011riemannian}
B.~Afsari.
\newblock Riemannian l p center of mass: Existence, uniqueness and convexity.
\newblock In {\em Proc. Amer. Math. Soc}, volume 139, pages 655--673, 2011.

\bibitem{bacciotti1999stability}
A.~Bacciotti and F.~Ceragioli.
\newblock Stability and stabilization of discontinuous systems and nonsmooth
  lyapunov functions.
\newblock {\em Control Optimisation and Calculus of Variations}, 4:361--376,
  1999.

\bibitem{bacciotti2005invariance}
A.~Bacciotti and L.~Mazzi.
\newblock An invariance principle for nonlinear switched systems.
\newblock {\em Systems \& Control Letters}, 54(11):1109--1119, 2005.

\bibitem{cheng2008}
D.~Cheng, J.~Wang, and X.~Hu.
\newblock An extension of lasalle's invariance principle and its application to
  multi-agent consensus.
\newblock {\em IEEE Transactions on Automatic Control}, 53(7):1765--1770, 2008.

\bibitem{clarke1975generalized}
F.~H. Clarke.
\newblock Generalized gradients and applications.
\newblock {\em Transactions of the American Mathematical Society},
  205:247--262, 1975.

\bibitem{clarke1990optimization}
F.H. Clarke.
\newblock {\em Optimization and nonsmooth analysis}, volume~5.
\newblock Siam, 1990.

\bibitem{garin2010survey}
F.~Garin and L.~Schenato.
\newblock A survey on distributed estimation and control applications using
  linear consensus algorithms.
\newblock In {\em Networked Control Systems}, pages 75--107. Springer, 2010.

\bibitem{Hartley2011}
J.~Hartley, R.~Trumpf and Y.~Da.
\newblock {Rotation averaging and weak convexity}.
\newblock {\em In Proceedings of the 19th International Symposium on
  Mathematical Theory of Networks and Systems (MTNS)}, pages 2435--2442, 2010.

\bibitem{hong2007}
Y.~Hong, L.~Gao, D.~Cheng, and J.~Hu.
\newblock Lyapunov-based approach to multiagent systems with switching jointly
  connected interconnection.
\newblock {\em IEEE Transactions on Automatic Control}, 52(5):943--948, 2007.

\bibitem{jadbabaie2003}
A.~Jadbabaie, J.~Lin, and A.~S. Morse.
\newblock Coordination of groups of mobile autonomous agents using nearest
  neighbor rules.
\newblock {\em IEEE Transactions on Automatic Control}, 48(6):988--1001, 2003.

\bibitem{khalil2002nonlinear}
H.~K. Khalil.
\newblock {\em Nonlinear systems}, volume Third Edition.
\newblock Prentice hall, 2002.

\bibitem{lin2007state}
Francis~B. Lin, Z. and M.~Maggiore.
\newblock State agreement for continuous-time coupled nonlinear systems.
\newblock {\em SIAM Journal on Control and Optimization}, 46(1):288--307, 2007.

\bibitem{ma2010necessary}
C-Q. Ma and J-F. Zhang.
\newblock Necessary and sufficient conditions for consensusability of linear
  multi-agent systems.
\newblock {\em Automatic Control, IEEE Transactions on}, 55(5):1263--1268,
  2010.

\bibitem{ma2004invitation}
Y.~Ma, S.~Soatto, J.~Kosecka, and S.~Sastry.
\newblock {\em {An invitation to 3-D vision}}.
\newblock Springer, 2004.

\bibitem{mesbahi2010graph}
M.~Mesbahi and M.~Egerstedt.
\newblock {\em Graph theoretic methods in multiagent networks}.
\newblock Princeton University Press, 2010.

\bibitem{montijano2011fast}
E.~Montijano, J.~I. Montijano, and C.~Sagues.
\newblock Fast distributed consensus with chebyshev polynomials.
\newblock In {\em American Control Conference (ACC), 2011}, pages 5450--5455.
  IEEE, 2011.

\bibitem{montijanoepipolar}
Eduardo Montijano, Johan Thunberg, Xiaoming Hu, and C~Sagues.
\newblock Epipolar visual servoing for multirobot distributed consensus.
\newblock {\em IEEE Transactions on Robotics}, pages 1--14.

\bibitem{moreau2004stability}
L.~Moreau.
\newblock Stability of continuous-time distributed consensus algorithms.
\newblock In {\em Decision and Control, 2004. CDC. 43rd IEEE Conference on},
  volume~4, pages 3998--4003. IEEE, 2004.

\bibitem{moreau2005stability}
L.~Moreau.
\newblock Stability of multiagent systems with time-dependent communication
  links.
\newblock {\em IEEE Transactions on Automatic Control}, 50(2):169--182, 2005.

\bibitem{olfati2007consensus}
R.~Olfati-Saber, J.A. Fax, and R.M. Murray.
\newblock Consensus and cooperation in networked multi-agent systems.
\newblock {\em Proceedings of the IEEE}, 95(1):215--233, 2007.

\bibitem{olfati2004consensus}
R.~Olfati-Saber and R.~M. Murray.
\newblock {Consensus problems in networks of agents with switching topology and
  time-delays}.
\newblock {\em Automatic Control, IEEE Transactions on}, 49(9):1520--1533,
  2004.

\bibitem{murray2003consensus}
R.~Olfati~Saber and R.M. Murray.
\newblock Consensus protocols for networks of dynamic agents.
\newblock In {\em Proceedings of the 2003 American Controls Conference}, 2003.

\bibitem{ren2005consensus}
W.~Ren and R.~W. Beard.
\newblock Consensus seeking in multiagent systems under dynamically changing
  interaction topologies.
\newblock {\em Automatic Control, IEEE Transactions on}, 50(5):655--661, 2005.

\bibitem{ren2005survey}
W.~Ren, R.~W. Beard, and E.M. Atkins.
\newblock A survey of consensus problems in multi-agent coordination.
\newblock In {\em American Control Conference, 2005. Proceedings of the 2005},
  pages 1859--1864. IEEE, 2005.

\bibitem{ren2008distributed}
W~Ren and R.W. Beard.
\newblock {\em Distributed consensus in multi-vehicle cooperative control:
  theory and applications}.
\newblock Springer, 2008.

\bibitem{reynolds1987flocks}
C.~W. Reynolds.
\newblock Flocks, herds and schools: A distributed behavioral model.
\newblock In {\em ACM SIGGRAPH Computer Graphics}, volume~21, pages 25--34.
  ACM, 1987.

\bibitem{rouche1977stability}
N.~Rouche, P.~Habets, M.~Laloy, and A-M. L.
\newblock {\em Stability theory by Liapunov's direct method}.
\newblock Springer-Verlag New York, 1977.

\bibitem{shi2009}
G.~Shi and Y.~Hong.
\newblock Global target aggregation and state agreement of nonlinear
  multi-agent systems with switching topologies.
\newblock {\em Automatica}, 45(5):1165--1175, 2009.

\bibitem{distthunberg}
J.~Thunberg, E.~Montijano, and X.~Hu.
\newblock Distributed attitude synchronization control.
\newblock In {\em 50th IEEE Conference on Decision and Control and European
  Control Conference}, pages 1962--1967. IEEE, 2011.

\bibitem{vicsek1995novel}
T.~Vicsek, A.~Czir{\'o}k, B-J. Eshel, I.~Cohen, and O.~Shochet.
\newblock Novel type of phase transition in a system of self-driven particles.
\newblock {\em Physical Review Letters}, 75(6):1226, 1995.

\bibitem{xiao2007distributed}
L.~Xiao, S.~Boyd, and S-J. Kim.
\newblock Distributed average consensus with least-mean-square deviation.
\newblock {\em Journal of Parallel and Distributed Computing}, 67(1):33--46,
  2007.

\bibitem{yoshizawa1966stability}
T.~Yoshizawa.
\newblock {\em Stability theory by Liapunov's second method}.
\newblock Mathematical Society of Japan (Tokyo), 1966.

\end{thebibliography}
\appendix
\end{document}